\documentclass[10pt, conference, compsocconf]{IEEEtran}

\usepackage{fancyhdr}
\usepackage[normalem]{ulem}

\usepackage[bookmarks=false,breaklinks=true,letterpaper=true,colorlinks,linkcolor=black,citecolor=blue,urlcolor=black]{hyperref}

\usepackage{epsfig}
\usepackage{graphicx}
\usepackage{epstopdf}
\usepackage{setspace} 
\usepackage{subfig}
\usepackage[ruled,vlined,linesnumbered]{algorithm2e}
\usepackage{url}
\usepackage{amsmath}
\usepackage{pifont}
\usepackage{fixltx2e}
\usepackage{anyfontsize}
\usepackage{kantlipsum}
\usepackage{multirow}
\usepackage[font=bf]{caption}
\usepackage{array}
\usepackage{comment}
\usepackage{mwe}
\let\labelindent\relax
\usepackage{enumitem}

\usepackage{tikz}

\newcommand\blfootnote[1]{%
  \begingroup
  \renewcommand\thefootnote{}\footnote{#1}%
  \addtocounter{footnote}{-1}%
  \endgroup
}

\newcommand{\mycomment}[1]{}
\newcommand{\ignore}[1]{}

\usepackage{color}
\usepackage{soul}
\usepackage{xspace}

\begin{document}
\title{FUSE: Fusing STT-MRAM into GPUs to Alleviate Off-Chip Memory Access Overheads}

\author{
{\large Jie Zhang$^{1}$, Myoungsoo Jung$^{1}$, and Mahmut Taylan Kandemir$^{2}$} \\
\vspace{-2pt}
       {\large \emph{Computer Architecture and Memory Systems Laboratory,}}\\
\vspace{-2pt}
       {\normalsize{Yonsei University$^{1}$, Pennsylvania State University$^{2}$}}\\
\vspace{-3pt}
	   {\large {http://camelab.org}}
       }

\maketitle

\begin{abstract}
In this work, we propose FUSE, a novel GPU cache system that integrates spin-transfer torque magnetic random-access memory (STT-MRAM) into the on-chip L1D cache. FUSE can minimize the number of outgoing memory accesses over the interconnection network of GPU's multiprocessors, which in turn can considerably improve the level of massive computing parallelism in GPUs. \blfootnote{This paper is accepted by and will be published at 25th IEEE International Symposium on High-Performance Computer Architecture (HPCA 2019). This document is presented to ensure timely dissemination of scholarly and technical work.}
Specifically, FUSE predicts a read-level of GPU memory accesses by extracting GPU runtime information and places write-once-read-multiple (WORM) data blocks into the STT-MRAM, while accommodating write-multiple data blocks over a small portion of SRAM in the L1D cache. To further reduce the off-chip memory accesses, FUSE also allows WORM data blocks to be allocated anywhere in the STT-MRAM by approximating the associativity with the limited number of tag comparators and I/O peripherals. Our evaluation results show that, in comparison to a traditional GPU cache, our proposed heterogeneous cache reduces the number of outgoing memory references by 32\% across the interconnection network, thereby improving the overall performance by 217\% and reducing energy cost by 53\%.

\end{abstract}


\IEEEpeerreviewmaketitle

\section{Introduction}
\label{sec:introduction}

Modern GPUs can achieve an outstanding performance with low power, which makes them very attractive for many non-graphics applications such as big-data analytics and scientific programs \cite{stuart2011multi,saecker2013big,chetlur2014cudnn,li2014large}, in addition to graphics applications. 

To properly utilize the massive computational parallelism, most GPUs employ multiple streaming multiprocessors (SMs) with large private register files. These large register files remove all the overheads imposed by task switches by accommodating multiple contexts along with many threads. However, owing to a large number of task contexts that each SM needs to keep track of, the private register files occupy a significant on-chip area. For example, Tesla P100, one of the recent NVIDIA GPUs \cite{pascalnvidia}, employs a 256KB register file per SM, which is 5$\times$ larger than the level one data (L1D) cache. In fact, these register files occupy 62\% of the total private memory area. The register files and L1D cache in many GPUs are integrated with the SMs on the same chip for active threads to hold their contexts and data. On the other hand, the SMs interface with off-chip memory to handle large I/O data for data-intensive applications. The off-chip DRAM is shared (as a global memory) by all the SMs through a high-speed interconnection network \cite{bakhoda2009analyzing}. Even though this off-chip memory architecture allows GPU kernels to process their data without a host-side software intervention, a reference to the off-chip memory requires going through all the hardware modules in the outgoing datapath of the GPU. 


\begin{figure}
\centering
\def\subfigcapskip{0pt}
\subfloat[Execution decomposition.]{\label{fig:small-read-perf}\rotatebox{0}{\includegraphics[width=0.49\linewidth]{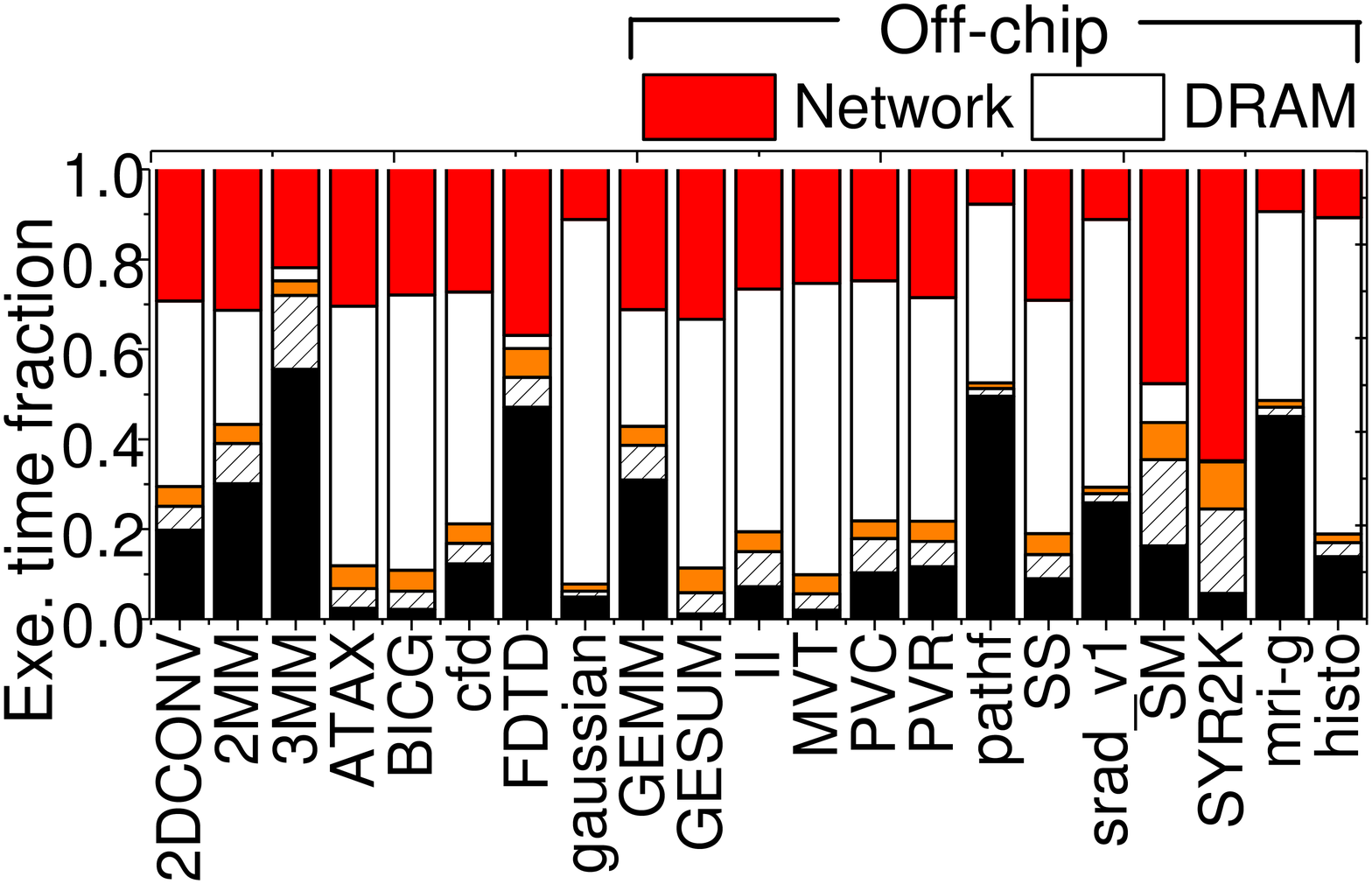}}}
\hspace{1pt}
\subfloat[Energy decomposition.]{\label{fig:small-read-utilidle}\rotatebox{0}{\includegraphics[width=0.49\linewidth]{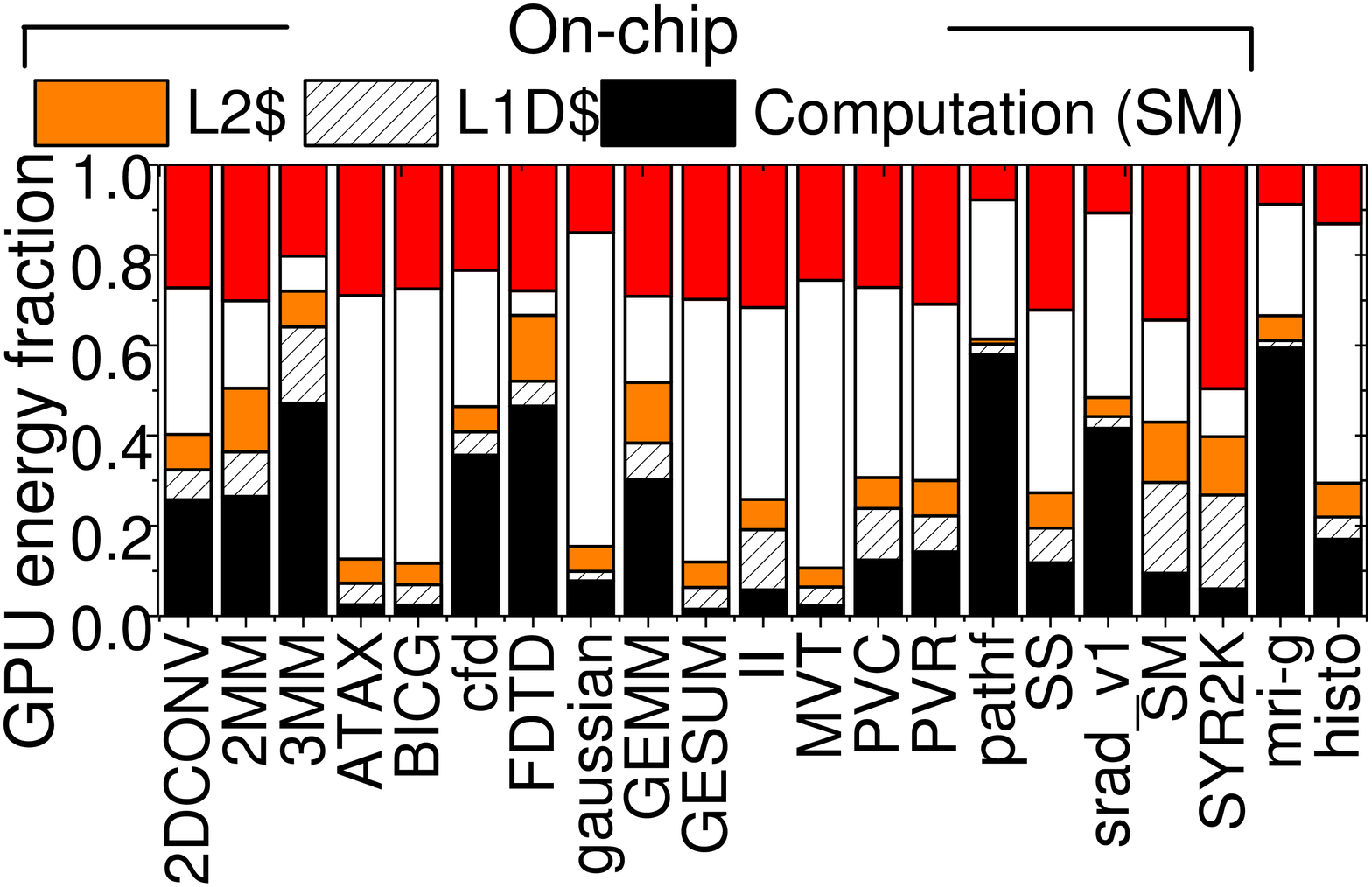}}}
\vspace{-5pt}
\caption{Overhead analysis for off-chip memory.\vspace{-10pt}}
\label{fig:motive}
\vspace{-15pt}
\end{figure}

Many large-scale GPU applications generate various irregular memory accesses, which need a relatively shorter response time. However, the DRAM employed by modern GPUs is designed for bringing out data with higher bandwidth using a wide I/O interface rather than shorter latency \cite{GDDR}. The long datapath between an SM and the DRAM also contributes to the latency of the off-chip memory accesses. Therefore, the off-chip access latency is much longer than the on-chip access latency.  
To be precise, we performed a simulation-based study, and the impact of the off-chip memory references on execution latency and energy consumption are shown in Figure \ref{fig:motive}. In this evaluation, we execute a diverse set of data-processing benchmarks \cite{pouchet2012polybench,che2009rodinia,stratton2012parboil,he2008mars} using a popular GPU model \cite{bakhoda2009analyzing}. The latency consumed by outgoing memory requests, on average, accounts for 75\% of the total GPU device execution time (cf. Figure \ref{fig:small-read-perf}). Similarly, the I/O service for off-chip data accesses takes up 71\% of the total GPU energy (cf. Figure \ref{fig:small-read-utilidle}). Even though there are many prior studies designed to optimize the on-chip memory accesses \cite{jia2014mrpb,chen2014adaptive} or parallelize kernel executions, one can observe from this analysis that such optimizations can only improve the overall performance of modern GPUs by at most 33\%. Consequently, the long latency and high energy incurred when accessing off-chip memory can be bottlenecks to deriving the full benefits of massive parallelism in GPUs.

Obviously, to maximize the performance of multi-threaded kernel executions, it is necessary for SMs to minimize the number of outgoing memory accesses on the GPU datapath. One simple solution can be to employ a larger L1D cache to accommodate the outgoing memory requests, thereby placing as many of them as possible close to the SMs. However, this simplistic solution is practically infeasible to achieve, since just enlarging L1D cache can severely reduce the on-chip space, which is already consumed by the GPU cores and register files. This in turn can reduce the maximum number of threads that could be executed in parallel, and can eventually deteriorate the throughput of GPU. 

Instead, the emerging concept of non-volatile memory (NVM) can be an option to overcome the on-chip area constraints and increase the storage capacity of L1D cache. Specifically, spin-transfer torque magnetic RAM (STT-MRAM) has great potential to replace SRAM on-chip modules. Thanks to its low-bit cost and small cell size, STT-MRAM can expand L1D cache capacity by around 4X (under the same area budget) \cite{liu2015efficient}. 

However, there are two main challenges to realizing this solution. First, constructing an L1D cache with pure STT-MRAM can considerably degrade the overall performance, irrespective of its superiority in terms of memory density. In fact, a write on STT-MRAM requires 5$\times$ longer operational time and consumes 3$\times$ higher power than an SRAM write, owing to the material-level characteristics of STT-MRAM \cite{huai2008spin}. Second, even with 4X bigger storage capacity that STT-MRAM delivers, the on-chip area of GPU is still insufficient to employ an ideal L1D cache, which could completely eliminate cache thrashing \cite{burtscher2012quantitative}.

In this work, we propose \emph{FUSE}, a novel heterogeneous GPU cache that integrates STT-MRAM into on-chip memory. The proposed FUSE can improve the overall GPU performance by significantly reducing the overheads imposed by off-chip memory accesses, while efficiently mitigating the write penalty of STT-MRAM. Our FUSE keeps a small portion of SRAM in the L1D cache to accommodate write-multiple data blocks, whereas it locates write-once-read-multiple (WORM) data blocks in the STT-MRAM that resides in its on-chip cache. FUSE can intelligently place each data block on either SRAM or STT-MRAM by employing a simple but effective online predictor that speculates a read-level for incoming memory references. 
We also introduce an associativity approximation logic that allocates WORM data blocks to any location in the STT-MRAM, while satisfying the tight on-chip area budget.
Our contributions in this work can be summarized as follows:

\noindent $\bullet$ \emph{Efficient heterogeneous GPU cache.} 
Our design exposes SRAM banks and STT-MRAM banks as one large on-chip storage pool to the GPU cores by enabling two memory modules fully operate in parallel. To this end, we apply non-blocking mechanism that internally employs a simple swap buffer and a tag queue.  The swap buffer frees SRAM from a hindrance to buffer the data that should be evicted to STT-MRAM, while the tag queue allows STT-MRAM to serve memory requests without any stall.
Overall, FUSE can improve performance by 217\%, on average, and reduce the energy by 53\%, in comparison to an SRAM-based L1D cache under diverse GPU application executions.

\noindent $\bullet$ \emph{Designing a smart data placement strategy.}
We observed that a majority of data blocks (around 90\%) experience WORM access patterns, which can render most bypassing approaches \cite{chen2014adaptive,li2015locality,li2015adaptive} in GPU impractical. Motivated by this, we design a read-level predictor that detects WORM data blocks by being aware of the execution behavior of kernels at runtime. We dynamically schedule the placement of data blocks associated with WORM into STT-MRAM, while forwarding the heavy writes to SRAM. By accurately tracking the history of GPU memory accesses, our data placement strategy can improve the GPU performance by 36\%, as compared to a conventional heterogeneous L1D cache.

\noindent $\bullet$ \emph{Approximated NVM cache.}
We propose an associativity approximation that fully utilizes STT-MRAM, while matching the area budget of a conventional set-associative cache. Our approximation logic places cache blocks in anywhere within STT-MRAM similar to a fully-associative cache, and serves the incoming requests as fast as a set associative cache would do by leveraging a counting bloom filter for rapid tag searching. The evaluation results reveal that this approximation technique can reduce the L1D cache misses by 23\%, on average, and improve the overall system performance by 172\%, compared to a set-associative NVM cache.

\ignore{
\noindent $\bullet$ \emph{Integrating STT-MRAM into GPU L1D cache.} 
Simply integrating STT-MRAM with SRAM as on-chip cache can stall the instruction pipeline of GPU due to STT-MRAM's write penalties. In this work, we introduce a swap buffer and a tag queue for heterogeneous cache architectures to enable non-blocking operations even with the write penalties. The swap buffer frees SRAM from a hindrance to buffer the data that should be evicted to STT-MRAM, while the tag queue allows STT-MRAM to serve incoming memory requests without a stall. This non-blocking mechanism also enables two memory modules to fully operate in parallel. Further, we design a bus arbitrator logic and a network that expose SRAM banks and STT-MRAM banks as one large on-chip storage pool to the GPU cores. 
Overall, the proposed FUSE can improve the performance by 217\%, on average, and reduce the system power by 53\%, in comparison to an SRAM-based L1D cache under diverse GPU application executions.

\noindent $\bullet$ \emph{Designing a smart data placement strategy.}
We observed that a majority of data blocks (around 90\%) experience WORM access patterns, which can render most bypassing approaches \cite{chen2014adaptive,li2015locality,li2015adaptive} in GPU impractical. This is because, a read miss for the data that has been bypassed to the underlying DRAM should be again placed into the L1D cache, which can bear the brunt of the write penalties of STT-MRAM. To address this challenge, we design a read-level predictor that detects WORM data blocks by being aware of GPU workload and runtime accesses. The proposed FUSE can dynamically schedule the placement of data blocks associated with WORM into STT-MRAM, while forwarding the heavy writes to SRAM. By effectively tracking the history of GPU memory accesses, the data placement strategy employed in our FUSE can improve the overall system performance by 36\%, as compared to a conventional heterogeneous L1D cache.

\noindent $\bullet$ \emph{Approximating NVM cache.}
Even though STT-MRAM offers a large storage capacity, the cache thrashing issue in GPU execution cannot be completely addressed, owing to poor cache utilization. Due to the limited area budget imposed by the on-chip architecture, it is not a feasible option to organize STT-MRAM as a fully-associative cache. Instead, we propose an associativity approximation logic that fully utilizes STT-MRAM, while matching the area budget of a conventional set-associative cache. Our approximation logic places cache blocks in anywhere within STT-MRAM similar to a fully-associative cache, and serves the incoming requests as fast as a set associative cache by leveraging a counting bloom filter for rapid tag searching. The evaluation results reveal that this approximation technique can reduce the L1D cache misses by 23\%, on average, and improve the overall system performance by 172\%, compared to a set-associative NVM cache.
}

\section{Background}
\label{sec:background}
\begin{figure}
\centering
\includegraphics[width=0.9\linewidth]{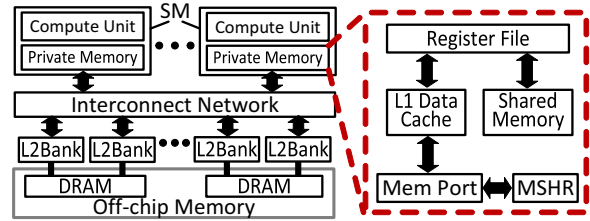}
\vspace{-10pt}
\caption{GPU baseline architecture.
\vspace{-5pt}}
\label{fig:baseline}
\vspace{-15pt}
\end{figure}



\subsection{GPU Architecture}

Figure \ref{fig:baseline} illustrates a typical GPU architecture that employs a dozen of SMs (i.e., simple in-order shader cores). Each SM consists of a compute unit and private on-chip memories. The compute unit contains 32 CUDA cores, which are organized as an in-order execution pipeline to process a group of 32 independent threads, referred to as a \emph{warp}, per GPU clock cycle \cite{wong2010demystifying}. To keep the pipeline busy, GPU schedules hundreds of threads within each SM and supports seamless context switches at the hardware level to eliminate the context switch penalty. On the other hand, the small private memories within each SM are used to cache instructions and data for fast response, while all input data are placed into a large shared off-chip memory (DRAM). The off-chip memory is connected to the SMs as a shared resource through an interconnection network and the data to be transferred are buffered in shared L2 caches \cite{chatterjee2014managing}. 
To access global data, the SMs need to go through all the components that exist in the data path to the off-chip memory behind the network.

\subsubsection{Private On-Chip Memory Components}

\noindent \textbf{\newline Registers.} Unlike CPUs that have a few registers, GPUs employ a large number of register files per SM to avoid the potential context switch overheads (without OS support). 
A GPU launches as many warps as possible in an SM, based on the availability of registers, to take advantage of massive parallelism without unnecessary register spilling.   
Since all the execution contexts for thread groups, such as local variables, always reside in the register files, the launched warps are never ``switched out". Even though the private register files allow the GPU to avoid all the procedures related to saving and restoring registers to/from memory, the registers occupy a significant portion of the on-chip area (e.g., 62\% of the private memory area in PASCAL GPU \cite{pascalnvidia}) to maintain a large number of contexts. 

\noindent \textbf{Shared memory.} The shared memory (also known as the scratchpad memory) is manually manipulated by programmers to load/store temporary data from/to registers for inter-warp communication. While the size of the shared memory can be reconfigured by reducing the size of the L1D cache, it should be determined at compile time. In addition, since all the shared memory references are statically determined and pinned, the off-chip memory traffic does not directly depend on its storage capacity or dynamic memory reference characteristics, but on the dynamic behaviors of applications. 

\begin{figure}
\centering
\def\subfigcapskip{0pt}
\subfloat[L1D miss rate.]{\label{fig:L1Dmiss-motiv}\rotatebox{0}{\includegraphics[width=0.49\linewidth]{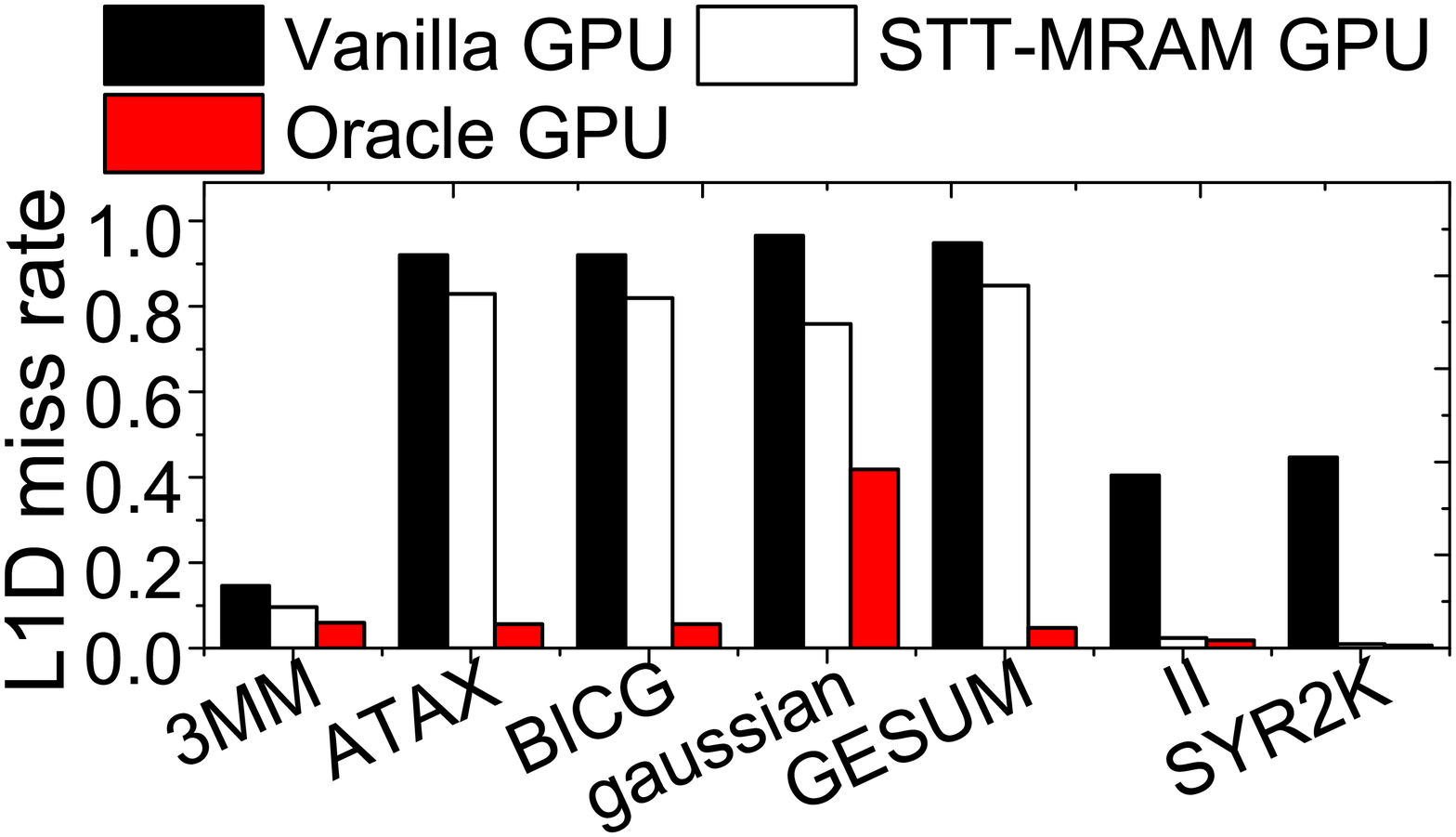}}}
\subfloat[IPC comparison.]{\label{fig:ipc-motiv}\rotatebox{0}{\includegraphics[width=0.49\linewidth]{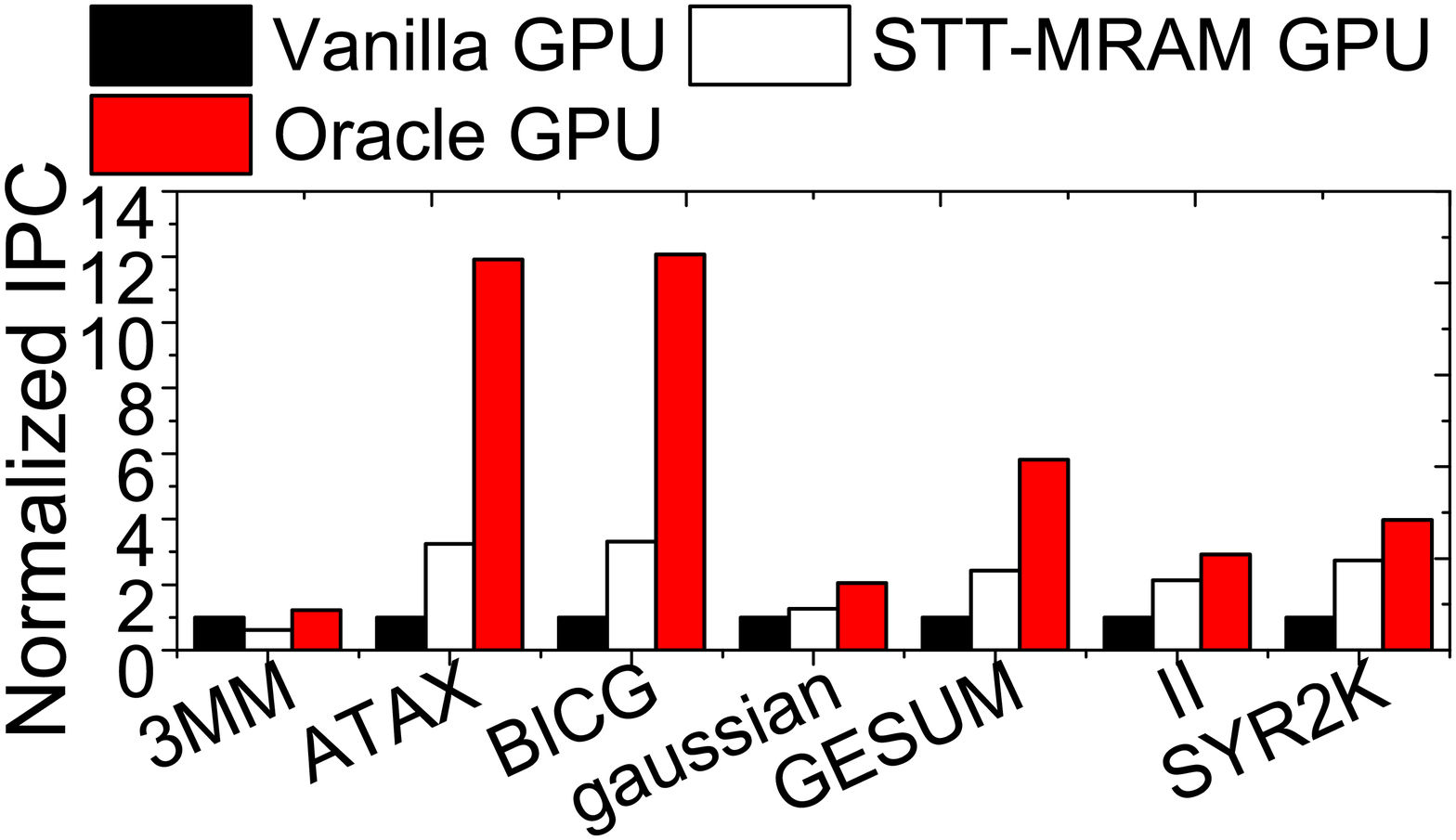}}}
\vspace{-10pt}
\caption{Performance for various L1D caches.\vspace{-15pt}}
\label{fig:moti}
\vspace{-10pt}
\end{figure}

\noindent \textbf{L1 data cache.} The L1D cache in GPU is, in practice, organized as a non-blocking SRAM-based cache \cite{chen1992reducing}. It is coupled with a parallel lower-level memory system by leveraging the miss status holding register (MSHR), which allows multiple misses to be merged with a primary miss (if it does not get issued to the next level of memory). 
While the L1D cache is designed to enable fast data access response and hide long off-chip memory access latency, the performance of a given GPU can significantly vary based on its storage capacity and cache organization. Specifically, modern GPUs execute thousands of threads, but their L1D caches are too small to accommodate all the data (i.e., up to 48KB per SM) \cite{wong2010demystifying}. This, in turn, leads to intensive competition among all active warps to grab cache lines and introduces cache thrashing, which replaces many data blocks before they are actually re-referred. To get a better insight, we evaluated the performance of seven representative memory-intensive workloads \cite{pouchet2012polybench,che2009rodinia} by using a popular GPU simulator \cite{bakhoda2009analyzing}, and the collected results are plotted in Figure \ref{fig:moti}. In this evaluation, the ``Vanilla GPU'' employs a small-sized L1D cache similar to that of GTX480 GPU \cite{glaskowsky2009nvidia}, whereas the ``Oracle GPU''  employs an ideal L1D cache that has enough capacity to avoid cache thrashing. 
The Oracle GPU reduces the L1D cache miss rate by 58\%, which in turn improves the overall performance by 6$\times$, compared to Vanilla GPU. 

\subsubsection{Shared L2 Cache and DRAM}

To support the data parallelism enabled by SMs, a GPU employs a large off-chip memory (i.e., DRAM) and L2 caches through a high-speed interconnection network. While the private on-chip memory modules are directly connected, the off-chip DRAM and L2 caches are connected to SM cores via an interconnect, which contributes to a significant round trip latency \cite{chen2014adaptive}. 
In addition, a memory reference, which misses in the private on-chip memory, experiences the latency of the L2 cache, which is 60 times longer than that of the L1D cache, as all the data are protected by ECC and its banks are shared by all SMs \cite{bakhoda2009analyzing}. If data does not exist in the L2 cache, the request is served by a global memory (DRAM). GPU DRAM is connected to the L2 cache through multiple memory channels. GPU DRAM also drastically increases the off-chip memory access latency, owing to i) its wide I/O interface that operates on a low frequency and ii) complicated I/O procedures to service a memory reference. The DRAM of a GPU employs a memory interface, that is 4X wider than a conventional DRAM, thereby enabling higher bandwidth with lower power and heat dispersal requirement. Even though this design concept can make GPU DRAM a high-performance module, the latency of GPU DRAM is slower than the conventional DRAM due to its complex internal timing. Further, DRAM has memory-specific operation delays such as precharges, which make DRAM much slower than on-chip storage. Modern DRAMs also queue all incoming references into multiple request queues for memory coalescing and reordering \cite{rixner2000memory}. Note that, as shown in Figure \ref{fig:motive}, the off-chip memory references account for 75\% of the total execution time and also consume 71\% of the total computing energy.

\begin{figure}
\centering
\includegraphics[width=1\linewidth]{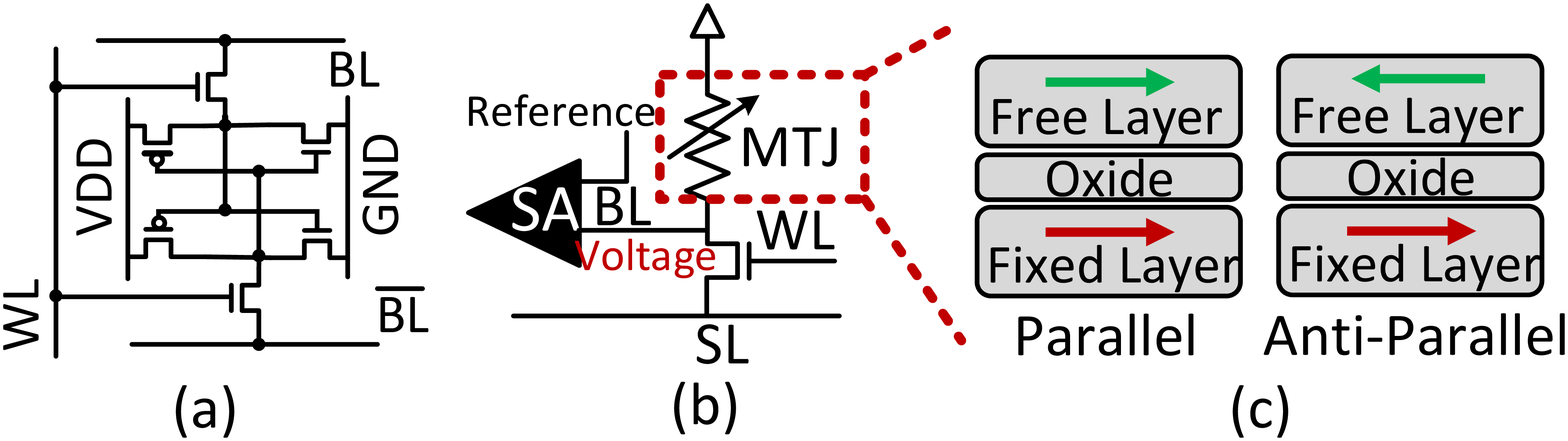}
\vspace{-20pt}
\caption{(a) SRAM memory cell, (b) STT-MRAM memory cell and (c) MTJ structure.
\vspace{-10pt}}
\label{fig:sttram}
\vspace{-10pt}
\end{figure}

\ignore{
\begin{table}
	\scriptsize
	\begin{center}
		\begin{tabular}{| c | c | }
		\hline
		Parameter & values \\
		\hline
		\hline
		Process technology & 22nm \\
		MTJ length, width, thickness & 45nm, 22nm, 3.2nm\\
		Operating temperature & 300K \\
		MTJ resistance (Anti-Parallel) & 3000 Ohm \\
		MTJ resistance (Parallel) & 1500 Ohm \\
		Switching probability $P_{sw}$ & 99.9\% \\
		Bohr's constant $\mu _{B}$ & $9.274\times 10^{-24}$ \\
		Electron charge,e & $1.602\times 10^{-16}$ \\
		Spin polarization,P & 0.75\\
		\hline
		\end{tabular}
	\end{center}
	\vspace{-15pt}
	\caption{STT-MRAM parameters for modelling.}
	\vspace{-5pt}
	\label{tbl:stt-para}
\end{table}
}

\subsection{STT-MRAM}
Figures \ref{fig:sttram}a and \ref{fig:sttram}b show the basic cell structures for SRAM and STT-MRAM, respectively. An SRAM cell comprises six transistors that hold 1-bit data in latch-like structures (cross-coupled inverter), which exhibits excellent performance in terms of both latency and dynamic energy for both read and write operations. However, the six-transistor structure requires $140F^2$ space \cite{itrs} and consumes a high static power due to sub-threshold and gate leakage currents. 
On the other hand, an STT-MRAM cell consists of a \emph{magnetic tunnel junction} (MTJ) and a single access transistor. This compact one-transistor-one-MTJ design leads to an STT-MRAM cell area of $36F^2$ \cite{sun2011multi, jin2014area, shihab2016couture}, which makes STT-MRAM about 4X denser compared to SRAM. 

\noindent \textbf{Architecture.}
As shown in Figure \ref{fig:sttram}c, an MTJ consists of two ferro-magnetic layers, a free layer and a fixed layer, which are isolated by oxide. The resistance of MTJ varies on the basis of the polarity of the free layer and the fixed layer. Specifically, if the two layers have the same polarity (parallel), MTJ exhibits low resistance, and it is typically considered as logic ``0". Otherwise, if the two layers have the opposite polarity (anti-parallel), MTJ has high resistance, and it is denoted as logic ``1". STT-MRAM writes data by using polarized electrons to directly torque the magnetic state of the free layer. As its writing process needs to physically rotate the MTJ free layer, an STT-MRAM write requires a long operation time and consumes a large amount of energy. On the other hand, the read operation is carried out by measuring the resistance of the cell. For this reason, STT-MRAM exhibits excellent read latency and energy, which can be comparable to SRAM. However, the long latency and high energy consumption on writes make it difficult for STT-MRAM to directly replace SRAM-based on-chip cache.

\noindent \textbf{Challenges.}  \label{sec:challenges}
Even though STT-MRAM can increase the storage capacity of the L1D cache by 4X, it does not completely address a cache thrashing issue.  
To be precise, we also simulate a GPU that employs STT-MRAM (``STT-MRAM GPU'') and compare its cache miss rate with the ``Vanilla GPU'' and the ``Oracle GPU'', which employ a popular L1D cache configuration and ideal configuration, respectively.
As shown in Figure \ref{fig:L1Dmiss-motiv}, the STT-MRAM GPU suffers from cache misses more than 39\%, on average, compared to the Oracle GPU. This is because irregular memory accesses originating from multiple GPU threads interfere with one another, which in turn leads to cache thrashing. In particular, STT-MRAM GPU reduces the L1D cache miss rate by less than 10\%, as compared to the Vanilla GPU in cases of ATAX, BICG, and GESUMMV. The reason why the STT-MRAM GPU has poor performance for these workloads is because of the write penalty exhibited by STT-MRAM as well as the memory access irregularity.

\begin{figure}
\centering
\includegraphics[width=1\linewidth]{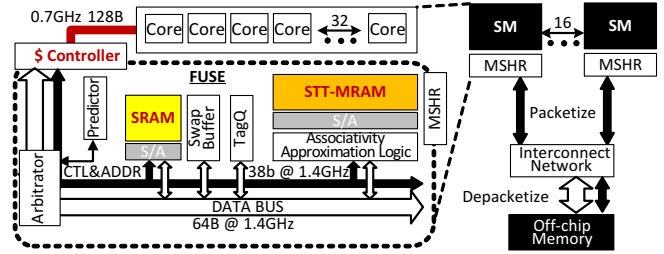}
\vspace{-15pt}
\caption{High-level view of the proposed FUSE.
\vspace{-10pt}}
\label{fig:overview}
\vspace{-15pt}
\end{figure}

\section{Fusing STT-MRAM With L1D Cache}
\label{sec:highlevelview}


\subsection{Heterogeneous Storage}

\noindent \textbf{Overall architecture.}
Figure \ref{fig:overview} illustrates a high-level view of our FUSE. SRAM and STT-MRAM are organized as separate cache banks, each employing their own individual data arrays and tag arrays, but operate as one large on-chip L1D cache. GPU cores in an SM can communicate with the on-chip cache controller over a 128B wide data bus, and this controller connects SRAM and STT-MRAM through a 64B wide data bus and independent 38-bit control lines. Since each thread (in a warp) can access a 4B data, executing a warp requires bringing in/out 128B data over a 700Mhz external bus. The 128B requests from the external interface can be coalesced and delivered by the 64B wide 1.4GHz internal data bus. We also employ 38-bit cache control lines, which are mainly used for delivering memory addresses (32 bits) and commands such as read and write signals (2 bits), byte-enable signals (3 bits), and acknowledge (1 bit) of our hybrid storage. All the misses from the heterogeneous cache are issued to the underlying MSHR that handles non-blocking memory operations and are delivered to the off-chip memory over an interconnection network.

\begin{figure}
\centering
\includegraphics[width=1\linewidth]{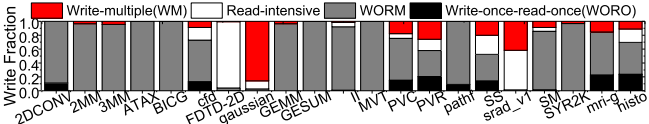}
\vspace{-20pt}
\caption{Read-level analysis.
\vspace{-10pt}}
\label{fig:read-level}
\vspace{-15pt}
\end{figure}

\noindent \textbf{Read-level analysis.}
The simplest way to utilize the memory heterogeneity is to place all incoming data into SRAM first and leverage STT-MRAM as a victim buffer of the SRAM \cite{zhang2016ross}. However, we observe that this simplisitic dynamic data placement strategy introduces excessive cache evictions from SRAM to STT-MRAM, due to the limited associativity and small storage capacity of the former. This in turn degrades the overall system performance by 63\%, on average, compared to the Oracle GPU. To address this, we analyzed the data block access patterns across diverse data-processing benchmarks \cite{pouchet2012polybench,che2009rodinia,stratton2012parboil,he2008mars}, and the results are shown in Figure \ref{fig:read-level}. For this evaluation, we traced all memory references, and based on read-level for each data block address, we categorized them into four types of requests: i) \emph{write-multiple} (WM), ii) \emph{read-intensive}, iii) \emph{write-once-read-multiple} (WORM), and iv) \emph{write-once-read-once} (WORO). WM indicates that the block has multiple updates, which may introduce many write hits on the cache. Read-intensive means the block is referred to by a few writes but by many reads. WORM is a purely read-oriented block, which refers to a block that only has a write and is never touched again over the entire execution of an application. Since a WORO data block is referred once during the execution, it is not necessary to put it in on-chip cache. There are three observations behind this analysis. First, since most data blocks are classified as WORM (80\% of total data blocks, on average), if there is a cache large enough to hold WORM data blocks, it can cut the off-chip memory accesses significantly. Second, there are several workloads (e.g., PVC, PVR, SS) that also include many WM data blocks. Due to the long STT-MRAM write latency, accommodating WM data blocks in STT-MRAM can deteriorate system performance. Third, as read-intensive data blocks have a few writes, SRAM is the best candidate to accommodate the data blocks. If SRAM size is not sufficient to hold the read-intensive data blocks, placing the data blocks in STT-MRAM can increase read cache hits and reduce off-chip memory accesses. Note that, if we bypass WORM accesses, it will introduce multiple writes for cache fills in a near future, which can also bear the brunt of the write penalty imposed by STT-MRAM. 

\noindent \textbf{Heterogeneity management.}
Motivated by the aforementioned observations, our FUSE employs a cache controller that \textit{speculates} the type of memory access based on GPU runtime information and can treat it accordingly. The cache controller refers to a program counter (PC) indexed table (inspired by \cite{khan2010sampling, ahn2014dasca, tian2015adaptive}), which records data access history of different PCs, to make a decision for the read-level of incoming memory references. Even though this is a simple history-based prediction, the speculation can be very effective in GPU, considering that i) each SM executes multiple thread groups that share all the same kernel instances, and ii) their memory references requested by the same PC generate access patterns similar to each other. Based on the prediction result, our cache controller places the corresponding cache line either at SRAM or STT-MRAM or evicts data to either STT-MRAM or L2 cache. Thus, only single data copy exists in either SRAM or STT-MRAM, which guarantees data consistency in hybrid L1D cache.
Since the cache controller may result in a wrong prediction, it is possible to place WORM data into SRAM, or to allocate WM data into STT-MRAM. 
In the former case, the cache controller migrates data from SRAM to STT-MRAM, which introduces a write and a read for STT-MRAM and SRAM, respectively. If the latter is a wrong prediction, it migrates data from STT-MRAM to SRAM right away and invalidates data in STT-MRAM, before SRAM serves the requests. 
The misprediction introduces extra cache operations, but the overall performance should be much better than pulling/pushing the data from/to the GPU DRAM. Our read-level predictor also exhibits a satisfactory prediction accuracy to reduce such overhead.

Migrating data from SRAM to STT-MRAM (on an internal cache replacement) can stall the instruction pipeline of the SM, due to long write latency of STT-MRAM. To avoid instruction stalls, we introduce a swap buffer, crossing different bank domain boundaries, which consists of multiple 128-byte data registers shared by both SRAM and STT-MRAM banks. The details of swap buffer will be discussed in Section \ref{sec:Arbitration}. 

\ignore{********* needs to be reconsidered
\noindent \textbf{Bank access parallelism.}
It is inefficient to block memory pipeline when STT-MRAM bank takes a long access time (tag search+data access) to respond incoming memory requests. To address this challenge, a non-blocking request queue is employed in our design. To be precise, the next memory request can be issued without waiting for STT-MRAM bank's response to current request. In this case, memory accesses to SRAM and STT-MRAM banks are not necessarily simultaneous and thus the future memory requests can directly visit fast SRAM bank. If the request hits in SRAM, it will be directly served; otherwise, it is pushed in request queue and waits for STT-MRAM bank's response.
}

\begin{figure}
\centering
\vspace{-10pt}
\subfloat[Associativity approximation logic.]{\rotatebox{0}{\includegraphics[width=0.75\linewidth]{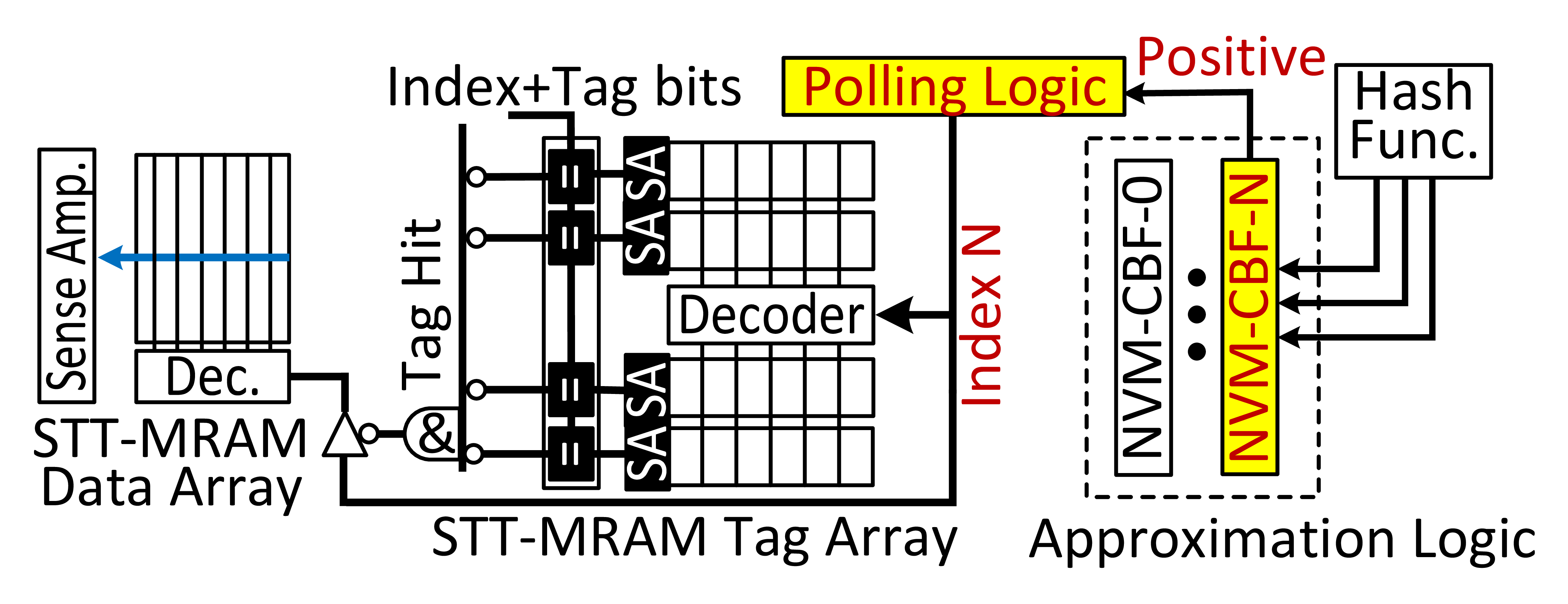}}}
\subfloat[Performance.]{\rotatebox{0}{\includegraphics[width=0.25\linewidth]{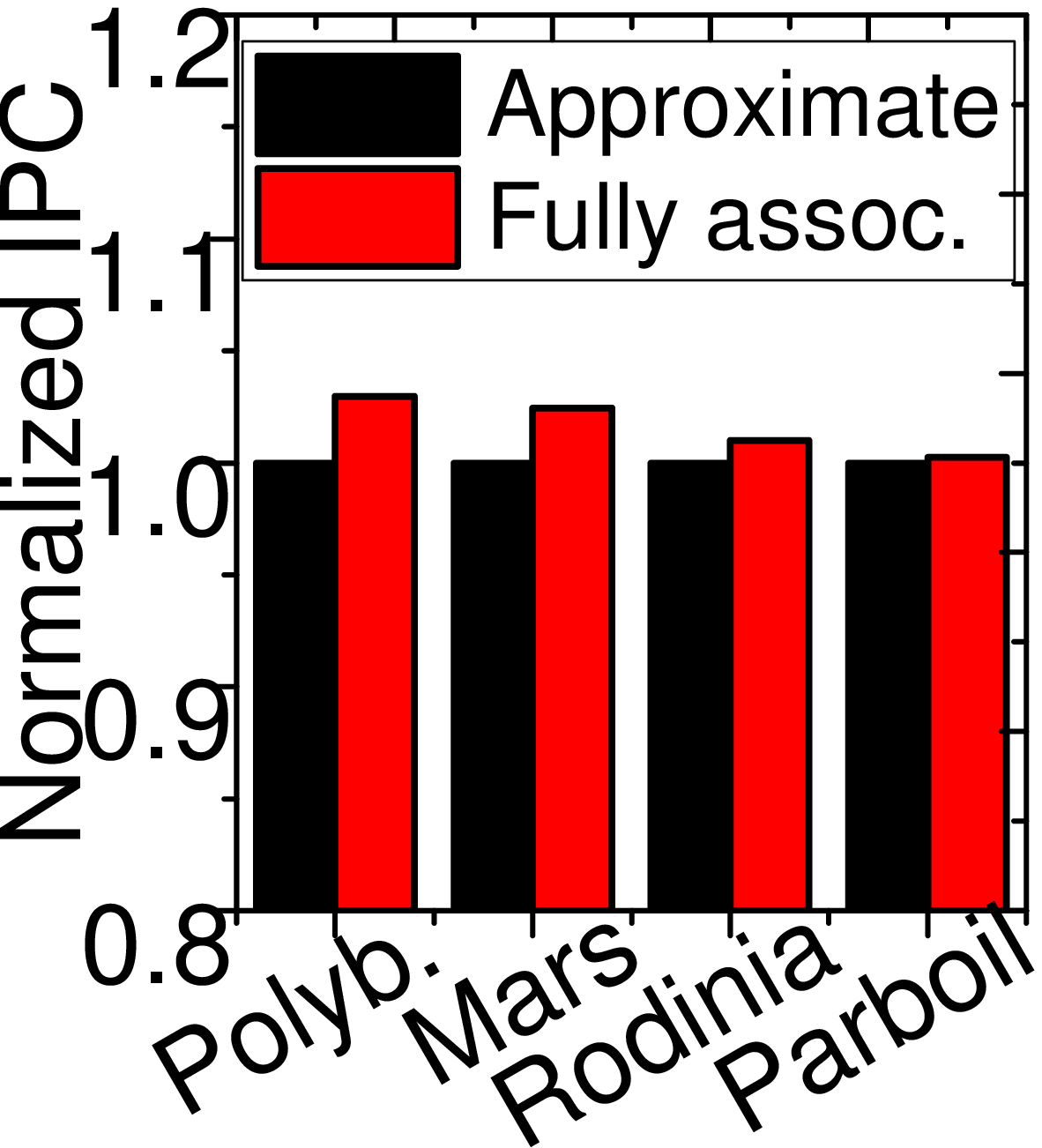}}}
\vspace{-5pt}
\caption{Performance for various L1D caches.\vspace{-10pt}}
\label{fig:tag_detail}
\vspace{-10pt}
\end{figure}

\subsection{Associativity Approximation} 
The theoretical feature size limit of STT-MRAM (around 36$F^2$) makes it impossible for FUSE to accommodate all WORM data blocks (including read-intensive) and eliminate GPU cache thrashing. Thus, one possible solution to minimize the number of outgoing memory requests is to place data into any entry of on-chip cache blocks like a fully-associative cache. However, the main hurdles that a fully-associative cache in GPU computing faces are its high hardware-level complexity and the operational costs of checking all tags in parallel. 
For example, 16KB fully-associative cache takes up 30.6 $\times$ more chip area and consumes 28.3 times more power than 4-way set associative cache \cite{wilton1996cacti}.
The main reason why the fully-associative cache occupies most on-chip area and consumes higher energy is the \textit{tag search}: the fully-associative cache needs as many comparators (that work in parallel) as its cache lines. 

Our FUSE employs an \textit{associativity approximation logic} to secure more WORM (and read-intensive) data by putting them in anywhere in STT-MRAM, while satisfying the tight on-chip area budget. Figure \ref{fig:tag_detail} shows details of the proposed associativity approximation logic. Generally speaking, the approximation logic serializes the tag comparison to some extent and uses a limited number of parallel comparators with polling logic -- in this work, there are four comparators. In each cycle, an index is inserted into the tag array decoder by the polling logic, and the tags of the same index are sensed out to compare them with the tag of incoming request. To avoid the long latency and energy overheads imposed by the serialized operations, the approximation logic has multiple \textit{Counting Bloom Filters} (CBFs) in front of the polling logic. It partitions the whole tag array as multiple data sets, each being mapped to a specific CBF, and examines whether target data resides in a data set by testing the corresponding CBF.  

While CBFs can effectively reduce tag search iterations by narrowing down search regions from the whole tag array to a few tag entries, the wrong results of CBFs can introduce multiple tag search iterations. In the worst case, the tag polling circuit generates all tag array indices to find out the target tag set. By carefully tuning the configurations of the approximation logic, we observe that tag search with the assistance of the approximation logic takes only 1 or 2 cycles across all the GPU workloads that we tested. Specifically, we compare the average performance (i.e., IPC) exhibited by our associativity approximation logic against that on a fully-associative cache. As shown in Figure \ref{fig:tag_detail}b, the performance difference between the two is under 2\% across all the benchmarks we tested. 
This is because, even though the tag search latency of approximation logic sometimes is longer than that of the fully-associative cache, the tag search does not stall the instruction execution pipeline by buffering the requests in tag queue. The design of tag queue will be explained shortly.

\section{Implementation Details}
\label{sec:implementation}

\subsection{Arbitration Logic}
\label{sec:Arbitration}

\ignore{
\begin{table}
\vspace{-10pt}
\centering
\resizebox{\textwidth}{!}{
\begin{tabular}{|c|c|c|c|}
\hline
SRAM status & STT-MRAM status & Approx. status & Arbitrator action \\ \hline
hit & X & X & SRAM->bus \\ \hline
miss & X & miss & request->L2 \\ \hline
miss & busy & hit & request->requstQ \\ \hline
X & hit & X & STT-MRAM->data bus \\ \hline
X & miss & X & request->requstQ \\ \hline
\end{tabular}}
\vspace{-15pt}
\caption{arbitration transition table.\vspace{-15pt}}
\label{tab:arb}
\end{table}
}

\begin{figure}
\centering
\includegraphics[width=1\linewidth]{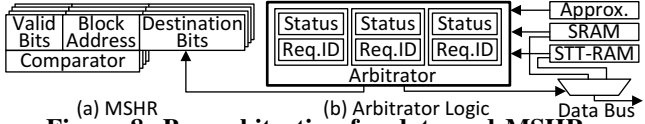}
\vspace{-25pt}
\caption{Bus arbitration for data and MSHR.
\vspace{-10pt}}
\label{fig:arbitrator}
\vspace{-5pt}
\end{figure}

\begin{figure}
\centering
\includegraphics[width=1\linewidth]{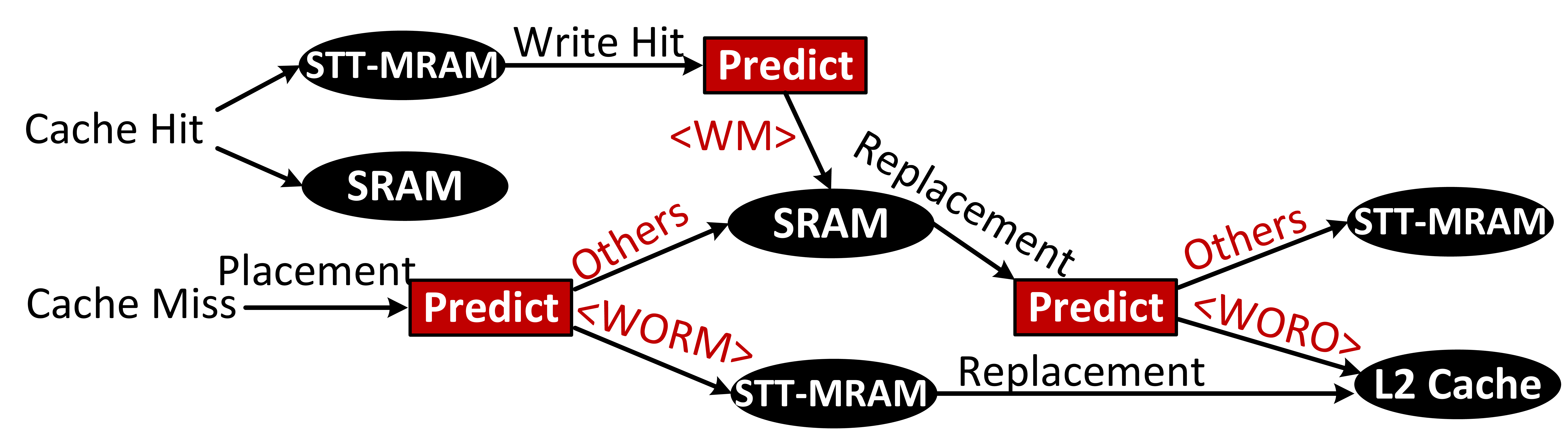}
\vspace{-20pt}
\caption{Decision tree of the arbitrator.
\vspace{-10pt}}
\vspace{-10pt}
\label{fig:dataflow}
\end{figure}


\noindent \textbf{Basic control.} 
Figures \ref{fig:arbitrator} and \ref{fig:dataflow} illustrate the control connection and decision tree of the FUSE arbitration logic, respectively. The main role of the arbitrator is to establish an appropriate datapath for a memory service by considering the status of each memory module and the decision tree. Thus, it has a set of status registers, each representing the result of SRAM, STT-MRAM, and associativity approximation logic. A status register can indicate either a hit, a miss, or a busy state. 
If a hit is observed in SRAM, the arbitrator terminates the tag search performed at STT-MRAM side and assigns the ownership of internal data bus to SRAM, so that the memory reference can be immediately serviced. If the hit is on STT-MRAM, the arbitrator checks the result of prediction and services the request from STT-MRAM. In the case of WM, the arbitrator migrates data from STT-MRAM to SRAM. In cases where a cache line is evicted from SRAM, the arbitrator checks the read-level predictor and evicts the cache line to L2 cache, if there is a WORO block. Otherwise, the data is evicted to STT-MRAM through the swap buffer and tag queue, as will be explained shortly. On the other hand, if a miss is observed, the cache controller needs to allocate a block (i.e., cache placement) by referring the result of prediction. In cases where a cache placement is required, it checks the read-level predictor and evicts the cache line to proper places. 
In addition, as WORM data blocks are placed when a cache miss is observed, the arbitrator needs to control the destination of its cache fill response. To this end, the arbitrator manipulates the miss status hold register (MSHR) before forwarding the memory reference to L2 cache. Specifically, as shown in Figure \ref{fig:arbitrator}a, a classic MSHR table \cite{farkas1997memory} contains a ``destination bits'' field, which indicates the destination address of the GPU module (e.g., instruction cache, data cache, or register files). 
We extend this field to identify the different cache lines that belong to SRAM or STT-RAM by specifying internal cache bank IDs.

It is to be noted that the execution of arbitration logic can harmonize with the procedure of cache access in a lockstep. In addition, our RTL synthesis experiment reveals the arbitration logic costs less than 1 ns, which is shorter than the single clock cycle of the L1D cache. Considering that the latency of arbitration logic is much shorter than that of cache access, we believe that our arbitration logic is not on the critical path and does not hurt the throughput of the L1D cache.

\noindent \textbf{Tag queue.}
STT-MRAM latency can vary based on the results of tag search and write penalty. This, in turn, may stall the instruction pipeline of SM to a certain extent. To make STT-MRAM a non-blocking module, we employ a tag queue, which can contain a command type, tag and index for multiple memory references (up to 16). This tag queue can improve bank-level parallelism within our hybrid storage by managing the memory requests that wait for a STT-MRAM service. Even though STT-MRAM is designed towards accommodating WORM data blocks, in cases where we have a write update for the data on STT-MRAM (due to a misprediction), our cache controller flushes the tag queue (if there is a waiting request) and handles the write. This is because the tag queue can contain only meta-information for the request, whereas a write is associated with 128 bytes data in GPUs. We observed that this flush situation exists for only 7\% of the total memory requests across all the workloads tested. 

\begin{figure}
\centering
\includegraphics[width=0.8\linewidth]{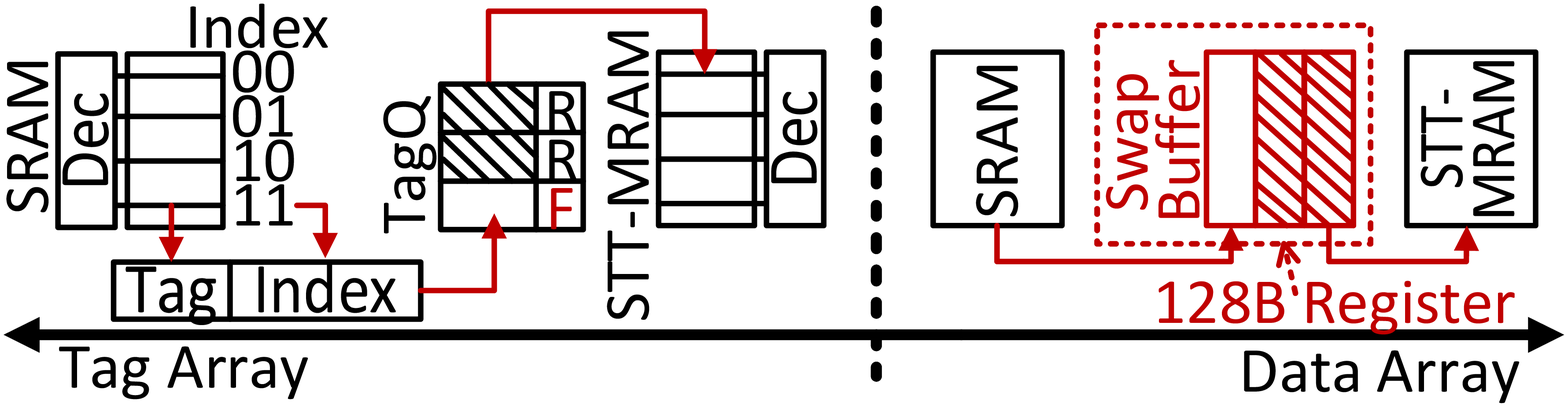}
\vspace{-12pt}
\caption{Architecture for non-blocking cache.
\vspace{-15pt}}
\label{fig:swapbuffer}
\vspace{-10pt}
\end{figure}

\noindent \textbf{Swap buffer.} 
We introduce a swap buffer to connect different bank domain boundaries and hide the write penalty in cases where there is a cache line eviction from SRAM. The swap buffer employs a few 128-byte data registers (up to 3 in this design), and each can buffer write data during STT-MRAM operations. When the cache controller evicts a line from SRAM, the data is immediately available from the swap buffer, but this approach requires snooping the buffer for coherence management.  Wu et al. \cite{wu2009hybrid} proposed a non-blocking swap buffer solution by adding extra comparators, data port, and data path to the swap buffer. Unfortunately, this solution does not fit well with our GPU L1D cache design due to high space cost. 
Instead, we address the coherence issue of non-blocking swap buffer by leveraging the aforementioned tag queue. As shown in Figure \ref{fig:swapbuffer}, when a data block is evicted from SRAM to STT-MRAM, the data block is directly migrated to the swap buffer. In the meantime, our cache controller rephrases the tag bits of the data block and index bits as a memory request and pushes it into the tag queue. Unlike a write on WORM update (i.e., wrong prediction), the target data exists in the swap buffer, which does not require flushing the tag queue. We mark this command with ``F'' in the tag queue, indicating a migration operation from swap buffer to STT-MRAM bank. Since the tag queue is managed as FIFO for only read and ``F" operations, matching a command and the corresponding data between the tag queue and swap buffer is straightforward. Therefore, FUSE can simply handle cache coherence in the heterogeneous storage \textit{without} snooping. 

\subsection{Read-Level Prediction}

\ignore{
\begin{table}
\centering
\scriptsize
\begin{tabular}{|c|c|c|c|c|c|}
\hline
\multicolumn{4}{|c|}{Sampler} & \multicolumn{2}{c|}{Prediction table} \\ \hline
\textbf{action} & used & mem addr & PC addr & counter & R/W \\ \hline
\textbf{insert} & reset & update & update & X & X \\ \hline
\textbf{read hit} & set & X & update & -1 & R \\ \hline
\textbf{write hit} & set & X & update & -1 & W \\ \hline
\textbf{evict} & X & X & X & !used bit?+1 & X \\ \hline
\end{tabular}
\vspace{-15pt}
\caption{Sampler status transition table\vspace{-15pt}}
\label{tab:sampler}
\end{table}
}

\begin{figure}
\centering
\includegraphics[width=1\linewidth]{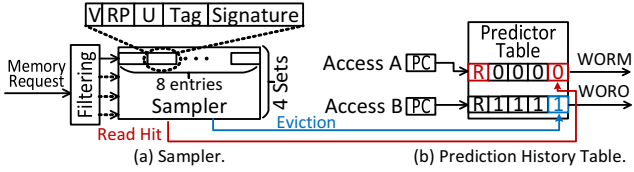}
\vspace{-20pt}
\caption{Read-level predictor.
\vspace{-10pt}}
\label{fig:predictor}
\vspace{-10pt}
\end{figure}

Our read-level predictor is aware of GPU memory access patterns by leveraging a PC-based predictor, which is similar to the one proposed in \cite{khan2010sampling, ahn2014dasca, tian2015adaptive}. Our prediction logic includes a memory request sampler and the prediction history table. Figure \ref{fig:predictor}a shows the detailed structure information for our predictor. 
The sampler consists of four sets, each containing eight entries, and operates like a 8-way set associative cache that employs LRU replacement policy. Each entry of the sampler stores information of the sampled memory request, including 1 valid bit (``V''), 1 used bit (``U''), 3 LRU control bits (``RP'), 15 partial memory address bits (``Tag''), and 9 partial PC address bits (``Signature''). 
In GPUs, warps from a given kernel are likely to execute a same set of instructions. Motivated by this, the sampler examines the memory requests generated from four representative warps (out of 48 warps) and places them into different sets in the sampler based on their warp number. The incoming requests should be compared with the ``Tag" bits in the sampler. If there is a miss, the new request is filled into the sampler entry and its ``U'' is reset to 0. Otherwise, a read/write hit in the entry sets ``U'' to 1.

The prediction history table is updated based on the result of sampler (e.g., hit or miss) and is referred by our arbitration logic to make a decision on read-level. Specifically, the history table is composed of an SRAM array that includes 512 entries indexed by ``Signature''. The table entry maintains a 1-bit R/W status and 4-bit counter. If there is a hit observed by the sampler, the counter of the history table referred by the corresponding ``Signature'' decreases. In case of a data eviction at the sampler, if ``U'' of the sampler is 0, the corresponding counter increases. 
Thus, if a counter is greater than a threshold ($unused_{th}$), the corresponding cache line is considered by arbitration logic as WORO. In this work, we observed the optimal $unused_{th}$ is fourteen across all the workloads tested. In contrast, if the counter is less than 1, the corresponding cache line is classified as either WORM or WM, based on whether the status bit is `R' or `W'. 
At the initial phase, the counter and status bit of a target entry are set to eight and `R', respectively. If the counter is in a range between 2 and $unused_{th}$ - 1, our arbitrator regards the target cache line as \emph{neutral}, which covers read-intensive block (cf. Figure \ref{fig:read-level}).


\begin{figure}
\centering
\includegraphics[width=1\linewidth]{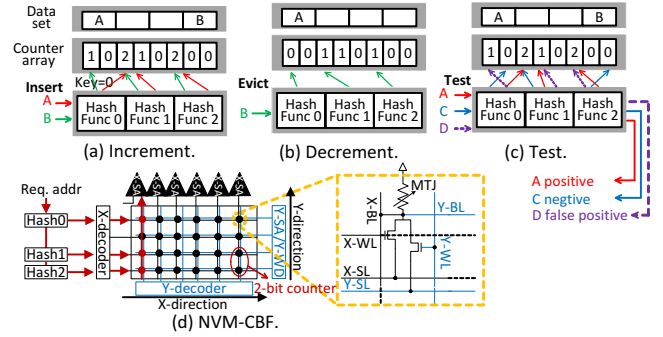}
\vspace{-20pt}
\caption{(a-c) Baseline counting bloom filter operations, (d) NVM-based CBF structure.
\vspace{-10pt}}
\label{fig:CBF}
\vspace{-10pt}
\end{figure}

\ignore{
\noindent \textbf{Integration.} As shown in Figure \ref{fig:predictor}c, when there is a cache miss, our arbitrator assigns SRAM to fetch neutral/WORO/WM cache lines and STT-MRAM to hold WORM cache lines. In addition FUSE cache controller utilize the predictor when it evicts a cache line of SRAM to specualte whether the target type is WORO. If the cache line contains WORO, the cache controller just sends it to L2 cache as they are just like deadwrites \cite{}. Otherwise, the target cache line will be evicted to STT-MRAM over the swap buffer as shown in Figure \ref{fig:predictor}b. Since the size of SRAM in our design much smaller than any conventional L1 cache, it is possible that SRAM cannot accommodate all WM data. Thus, in this worse case, a WM cache block can be temporally stored into STT-MRAM. If a write cache hit regarding the WM cache block is observed in STT-MRAM and there is a room at SRAM, the cache controller immediately migrates it back to SRAM, as in Figure \ref{fig:predictor}d.  This can be done sinc the predictor still keeps the read-level information based on the corresponding PC not memory address. 
}

\begin{table*}[]
\centering
\resizebox{\textwidth}{!}{
\begin{tabular}{|l|c||l|c|c|c|c|c|c|}
\hline
\multicolumn{2}{|l||}{\textbf{General Configuration}} &  & \multicolumn{1}{l|}{\textbf{L1-SRAM}} & \multicolumn{1}{l|}{\textbf{By-NVM}} & \multicolumn{1}{l|}{\textbf{Hybrid}} & \multicolumn{1}{l|}{\textbf{Base-FUSE}} & \multicolumn{1}{l|}{\textbf{FA-FUSE}} & \multicolumn{1}{l|}{\textbf{Dy-FUSE}} \\ \hline
\textbf{SMs/Request Q./swap buffer entries} & 15/16/3 & \textbf{SRAM/STT-MRAM size (KB)} & 32/0 & 0/128 & 16/64 & 16/64 & 16/64 & 16/64 \\ \hline
\textbf{CBF \#/hash function \#} & 128/3 & \textbf{SRAM/STT-MRAM leakage Power (mW)} & 58/0 & 0/2.8 & 36/2.6 & 36/2.6 & 36/2.4 & 36/2.4 \\ \hline
\textbf{Threads/Warp, Warps/SM, CTAs/SM} & 32, 48, 8 & \textbf{SRAM Set/Associativity} & 64/4 & 0/0 & 64/2 & 64/2 & 64/2 & 64/2 \\ \hline
\textbf{L1 SRAM/STT-MRAM latency} & 1/1 (R), 1/5-cycle (W) & \textbf{SRAM Sense Amplifier/Comparator} & 4/4 & 0/0 & 2/2 & 2/2 & 2/2 & 2/2 \\ \hline
\textbf{L2 cache size/sets/assoc./latency} & 786KB/64/8/1-cycle & \textbf{SRAM Read/Write Energy (nJ/access)} & 0.15/0.12 & 0 & 0.09/0.07 & 0.09/0.07 & 0.09/0.07 & 0.09/0.07 \\ \hline
\textbf{sampler assoc./sets} & 8/4 & \textbf{STT-MRAM Set/Associativity} & 0/0 & 256/4 & 256/2 & 256/2 & 1/512 & 1/512 \\ \hline
\textbf{Pred. history table entries/thres.} & 1024/14/1 & \textbf{STT-MRAM Sense Amplifier/Comparator} & 0/0 & 4/4 & 2/2 & 2/2 & 1/4 & 1/4 \\ \hline
\textbf{DRAM channels/tCL/tRCD/tRAS} & 6/12/12/28 & \textbf{STT-MRAM Read/Write Energy (nJ/access)} & 0/0 & 1.2/2.9 & 0.26/2.4 & 0.26/2.4 & 0.26/2.4 & 0.26/2.4 \\ \hline
\end{tabular}}
\vspace{-5 pt}
\caption{GPU simulation configuration.\label{tab:system_config}\vspace{-10 pt}}
\vspace{-5 pt}
\end{table*}

\begin{table}
\centering
\resizebox{\columnwidth}{!}{
\begin{tabular}{|c|c|c|c||c|c|c|c|}
\hline
\textbf{App} & \textbf{APKI} & \textbf{Bypass ratio} & \textbf{Suite} & \textbf{App} & \textbf{APKI} & \textbf{Bypass ratio} & \textbf{Suite} \\ \hline
2DCONV &9 & 0.26 & \cite{pouchet2012polybench} & 2MM &10 & 0.6 & \cite{pouchet2012polybench}  \\ \hline
gaussian &8.5 & 0.36 & \cite{stratton2012parboil} & 3MM &10 & 0.49 & \cite{pouchet2012polybench} \\ \hline
srad\_v1 &3.5 & 0.38 & \cite{che2009rodinia} & GEMM &136 & 0.61 & \cite{pouchet2012polybench} \\ \hline
ATAX &64 & 0.9 & \cite{pouchet2012polybench} & SYR2K &108 & 0.02 & \cite{pouchet2012polybench} \\ \hline
BICG &64 & 0.9  & \cite{pouchet2012polybench} & histo &9.6 & 0.63 & \cite{stratton2012parboil} \\ \hline
GESUMMV &12  & 0.96 & \cite{pouchet2012polybench} & mri-g &3.3 & 0.13 & \cite{stratton2012parboil} \\ \hline
FDTD-2D &18 & 0.27 & \cite{pouchet2012polybench} & pathf &1.2 & 0.92 & \cite{che2009rodinia} \\ \hline
MVT &64 & 0.91 & \cite{pouchet2012polybench}  & cfd &4.5 & 0.81 & \cite{che2009rodinia} \\ \hline
II &77 & 0.54 & \cite{he2008mars} & PVC &37 & 0.18 & \cite{he2008mars} \\ \hline
SS &30 & 0.80 & \cite{he2008mars} & SM &140 & 0.02 & \cite{he2008mars} \\ \hline
PVR &14 & 0.33 & \cite{he2008mars} & & & & \\ \hline 
\end{tabular}}
\vspace{-5 pt}
\caption{GPU benchmarks. \vspace{-5 pt}}
\vspace{-15 pt}
\label{tab:benchmark}
\end{table}

\subsection{NVM-based Counting Bloom Filter}

\noindent \textbf{Baseline bloom filter.}  
A counting bloom filter (CBF) is a space-efficient data structure, that can be used to test whether an element is a member of a specific set \cite{bonomi2006improved}. Figure \ref{fig:CBF} shows the CBF logic structure, which consists of multiple hash functions and a 1D counter array. The target data for a test is converted to several keys by hashing functions. These keys are then used as an address offset to access the counter array. If data is inserted in a data set, hash functions generate keys from the data (cf. Figure \ref{fig:CBF}a); these keys are used for locating specific counters in the counter array, and it then increments the counters by 1.  All these procedures are referred to as \emph{increment}. In contrast, in cases where data is evicted from the data set (cf. Figure \ref{fig:CBF}b), the counters corresponding to its keys decrement by 1, referred to as \emph{decrement}. Figure \ref{fig:CBF}c shows how to examine if data exists in the data set. In this test, if one of the counters contains value `0', it means that data does not exist in its data set, and is called ``negative''. Otherwise, it \textit{probably} exists in the data set, which is referred to as ``positive''. If the values in the checked counters are modified by other stored data, this results in false prediction, called as ``false positive''.

\noindent \textbf{Hardware support and modification.} 
The hardware structure is scalable, as our associativity approximation utilizes multiple CBFs. However, each baseline CBF needs multiple latches and transistors, which may occupy too much area of SRAM and violate L1D cache area constraints. Thus, we modify the baseline to use a separate STT-MRAM rather than SRAM in our heterogeneous storage, called NVM-CBF. As shown in Figure \ref{fig:CBF}d, it integrates the baseline CBFs in a 2D array structure by vertically locating a 2-bit counter array in the 2D MTJ island so that all counter arrays can share peripherals and I/O circuits.   
X-decoder simultaneously activates multiple sets of target rows associated with hashed keys, and all data stored in the arrays are compared to a reference voltage between the voltage of completely-zero or non-zero, and sensed out through a read port to distinguish between positive or negative. Owing to parallel operations of NVM-CBF, a test can be completed within the cycles of a single STT-MRAM read. 
For the increment and decrement operations, a Y-decoder enables a specific column based on the target CBF ID and updates the revised counters through a write port. We configure an X-decoder, X-WL (wordline), X-SL (sourceline), and X-BL (bitline) to transfer the voltage of an activated storage cell to X-SA (sensing amplifier), while using a Y-decoder, Y-WL, Y-SL, and Y-BL to connect Y-SA/Y-WD (sense amplifier/write driver) to the target storage core. We verify the NVM-CBF design by prototyping the circuit and simulating the read/program operations in Cadence \cite{simulator2005cadence}.

Since STT-MRAM has mostly WORM data for better data reuse rather than going through the entire GPU memory hierarchy, the operation type of most accesses for NVM-CBFs is ``test". 
Our experiments with CACTI \cite{wilton1996cacti} indicates that ``test" procedure costs 591 ps, which is much shorter than a single cache cycle. Even in cases of increment/decrement on NVM-CBF, its latency can be overlapped with that of the corresponding STT-MRAM data array write operation. The number of hash functions, the number of CBFs, the length of the counter array, and the size of each counter can significantly impact CBF's performance. We tune these four parameters based on the observation, and achieve best performance with 3 hash functions and 128 CBFs, each having 16 2-bit counters. This takes 512B area in total, making NVM-CBF small enough. 

\section{Evaluation}
\label{sec:evaluation}

\noindent \textbf{Simulation methodology.}
We implemented FUSE on top of GPGPU-Sim 3.2.2 \cite{bakhoda2009analyzing}. 
We used 15 SMs with 32KB register files per SM, which can execute up to 48 warps per SM and total 1536 threads per SM. The interconnection network is configured as a butterfly topology with 27 nodes (i.e., 15 SMs and 12 L2 cache banks). Our simulator employs 6 DRAM channels, each connecting to 2 L2 cache banks. We adopt the latency models and replacement policies of L1D cache, L2 cache and off-chip DRAM from GPGPU-Sim, which has been verified with real GPU devices \cite{wittenbrink2011fermi,keplernvidia}. As the circuit complexity of LRU is not affordable in a full-associative cache, in our experiments we employ FIFO as the replacement policy of STT-MRAM bank. Other popular low-cost cache replacement policies can also be integrated in the STT-MRAM bank \cite{kedzierski2010adapting}. We used GPUWattch \cite{leng2013gpuwattch} to evaluate the power consumption of the simulated GPU system, while leveraging CACTI 6.5 \cite{wilton1996cacti} and NVsim \cite{dong2014nvsim} to estimate the latency and power values of STT-MRAM bank as well as peripheral circuits. The detailed energy models of an SRAM bank and an STT-MRAM bank are given in Table \ref{tab:system_config}.

\noindent \textbf{L1D configurations.}
We evaluate the performances of seven different L1Ds, and their important characteristics are given in Table \ref{tab:system_config}. While \texttt{L1-SRAM} is configured as a 4-way set-associative SRAM-based on-chip cache, \texttt{FA-SRAM} reconstructs \texttt{L1-SRAM} as a fully-associative cache. \texttt{By-NVM} is implemented with pure STT-MRAM that integrates deadwrite prediction \cite{ahn2014dasca} to avoid frequent STT-MRAM writes. \texttt{By-NVM} bypasses \emph{deadwrite}, which is written once but not re-referenced before getting evicted from the cache. \texttt{Hybrid} integrates 2-way associative SRAM bank and 2-way associative STT-MRAM bank within the same chip. Since \texttt{Hybrid} does not implement swap buffers or tag queue, any write on STT-MRAM will result in a long L1D cache stall. \texttt{Base-FUSE} integrates swap buffers and tag queue into \texttt{Hybrid}. The difference between \texttt{Base-FUSE} and \texttt{FA-FUSE} is that \texttt{FA-FUSE} configures an STT-MRAM bank as approximate fully-associative cache. In addition, \texttt{Dy-FUSE} optimizes \texttt{FA-FUSE} by employing a read-level predictor to intelligently place cache blocks in either SRAM bank or STT-MRAM bank. All above caches are constructed with the same area budget that \texttt{L1-SRAM} uses.
   
\noindent \textbf{Workloads.} We evaluate a large collection of workloads from PolyBench \cite{pouchet2012polybench}, Rodinia \cite{che2009rodinia}, Parboil \cite{stratton2012parboil}, and Mars \cite{he2008mars} benchmarks. While PolyBench is a set of benchmarks that contain static control parts, and the workloads we tested are mainly related for polyhedral computation. Rodinia considers bioinformatics, data mining and classical algorithm, and Mars is a kind of MapReduce framework optimized GPU that includes inverted indexing and page view count. Lastly, Parboil is a collection of different scientific and commercial kernels including image processing, biomolecular simulation, fluid dynamics, and astronomy. Considering the space limit, we first evaluate all the workloads, exclude the results, which exhibit similar performance behavior each other, and select 21 representative workloads among them. These 21 workloads exhibit different memory access patterns in terms of memory regularity (or lack of it) and read/write intensiveness. 
We give brief descriptions, access per Kilo-instruction (APKI), and bypass ratio for By-NVM in Table \ref{tab:benchmark}. To perform our evaluations with fidelity, all workloads we tested have more than one billion instructions.

\begin{figure}
\centering
\includegraphics[width=1\linewidth]{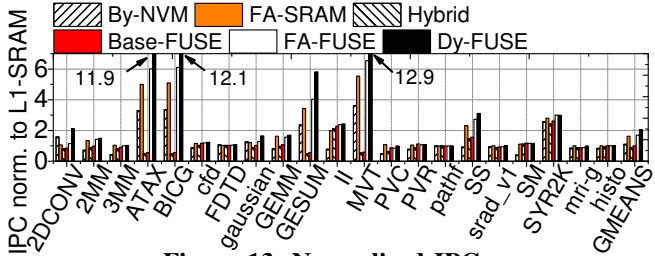}
\vspace{-20pt}
\caption{Normalized IPC.
\vspace{-10pt}
}
\vspace{-15pt}
\label{fig:ipc_002}
\end{figure}

\subsection{Overall Performance Analysis}
\noindent \textbf{Instructions per cycle (IPC).} 
Figure \ref{fig:ipc_002} shows the IPC values of the seven L1D configurations evaluated with our workloads, \emph{normalized} to \texttt{L1-SRAM}.
We note that \texttt{FA-SRAM} and \texttt{By-NVM} improve the performance by 63\% and 41\%, respectively, in regular read-intensive workloads (e.g., \emph{2DCONV} and \emph{gaussian}) and irregular workloads (e.g., \emph{ATAX} and \emph{GESUM}), compared to \texttt{L1-SRAM}. This is because both \texttt{FA-SRAM} and \texttt{By-NVM} can accommodate more memory requests by taking advantage of full associativity and leveraging the large STT-MRAM capacity, respectively. However, \texttt{FA-SRAM} is not a realistic option, due to its complex circuit design and high area overhead, while \texttt{By-NVM} suffers from long STT-MRAM write latency and long off-chip access delay caused by deadwrite eviction. The performance of \texttt{By-NVM} is 43\% worse than \texttt{L1-SRAM} in the write-intensive workloads such as \emph{2MM} and \emph{3MM}). 
\texttt{Hybrid} can reduce the STT-MRAM write penalty by fusing SRAM and STT-MRAM in the on-chip cache. However, without full optimization, \texttt{Hybrid} degrades the performance by 23\%, compared to \texttt{L1-SRAM}, in all the workloads.  
Since \texttt{Base-FUSE}, \texttt{FA-FUSE}, and \texttt{Dy-FUSE} optimize \texttt{Hybrid} from different aspects, they achieve, on average, 11\%, 206\%, and 398\% higher IPC than \texttt{Hybrid}, respectively.
This is because, instead of stalling for STT-MRAM writes, \texttt{Base-FUSE} enables an SRAM bank to serve future data requests and buffers the future STT-MRAM requests and data in the tag array and the swap buffer, respectively. Compared to \texttt{Base-FUSE}, \texttt{FA-FUSE} implements a realistic fully-associative STT-MRAM cache to accommodate as many memory requests as possible, and it improves the IPC by 4.1x in irregular workloads (e.g., \emph{ATAX}, and \emph{GEMM}).
\texttt{Dy-FUSE} performs better than \texttt{FA-FUSE} by 23.7\%, on average, across all the workloads. This is because \texttt{Dy-FUSE} can predict the access patterns of data requests and reduce the number of writes in STT-MRAM. Although \texttt{By-NVM} can allocate more STT-MRAM space than \texttt{Dy-FUSE}, its effectiveness in reducing L1D cache miss rate is limited. As a result, \texttt{Dy-FUSE} improves performance by 101\%, compared to \texttt{By-NVM}.

\begin{figure}
\centering
\includegraphics[width=1\linewidth]{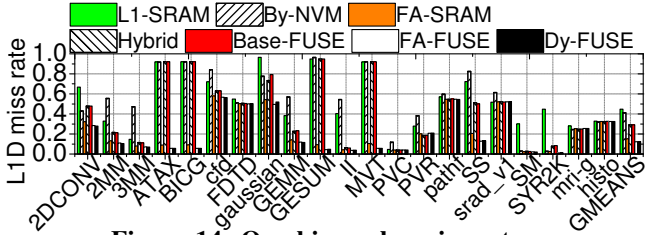}
\vspace{-20pt}
\caption{On-chip cache miss rate.
\vspace{-10pt}
}
\label{fig:L1_mr}
\end{figure}

\noindent \textbf{Cache miss rate.}
Figure \ref{fig:L1_mr} plots the L1D cache miss rates for different L1D cache options across different workloads. Among all tested L1D cache options, \texttt{L1-SRAM} has the highest cache miss rate in most workloads, due to the limited on-chip cache resources. 
\texttt{FA-SRAM} reduce the cache miss rate by 29\%, on average, compare to \texttt{L1-SRAM}. This is because, the full associativity of \texttt{FA-SRAM} can effectively reduce the L1D cache conflict misses.
Since \texttt{By-NVM} has 3x larger capacity than \texttt{L1-SRAM} (and consequently it can accommodate more cache requests), it exhibits a 24\% lower L1D miss rate than \texttt{L1-SRAM} in read-intensive workloads (e.g., \emph{2DCONV}). However, the on-chip cache miss rate of \texttt{By-NVM} depends on how many data requests have been bypassed to off-chip memories. For example, only 24\% of data requests are bypassed in \emph{2DCONV}, as indicated in Table \ref{tab:benchmark}. On the other hand, \emph{2MM} and \emph{3MM}, which contain more than 40\% of write requests, bypass more than 80\% of data requests. As a result, \texttt{By-NVM} has 28\% higher L1D cache miss rate than \texttt{L1-SRAM} in \emph{2MM} and \emph{3MM}, on average. Since workloads with irregular access patterns (e.g., \emph{ATAX} and \emph{BICG}) cannot reuse data blocks due to cache thrashing, such non-reused data blocks are regarded as deadwrite in \texttt{By-NVM}, which in turn generates a high bypassing ratio. Although \textit{By-NVM} cannot reduce miss rates in these workloads, the data requests are directly bypassed to the underlying L2 cache without any pause to obtain a free cache line in L1D cache, and this in turn results in higher performance improvement (cf., Figure \ref{fig:ipc_002}). 
\texttt{Hybrid} and all FUSE caches exhibit 21.6\% lower on-chip cache miss rate than that of \texttt{L1-SRAM}, on average, across all workloads. This is because these cache configurations increase the total on-chip cache size to better serve data requests from a massive number of active warps. Since \texttt{Base-FUSE} and \texttt{Hybrid} have similar L1D cache internal architectures, except for swap buffers and tag queue, they achieve the same on-chip cache miss rate, which is 10.5\% lower than that of \texttt{L1-SRAM}. On the other hand, \texttt{FA-FUSE} can reduce the on-chip cache miss rate by up to 86\% in irregular workloads \emph{2MM}, \emph{3MM}, \emph{ATAX}, \emph{BICG}, \emph{GEMM}, \emph{GESUM}, \emph{II}, \emph{MVT}, \emph{PVC}, \emph{SS}, \emph{SM}, and \emph{SYR2K}. This is because configuring STT-MRAM as a fully-associative cache can successfully reduce cache conflicts and accommodate as many irregular cache access patterns as possible. 
Lastly, since our read-level predictor is designed to detect the correct cache bank to put data blocks rather than modifying the placement/replacement policy for each cache bank, there is little difference between \texttt{FA-FUSE} and \texttt{Dy-FUSE} as far as cache miss rate is concerned.

%

\begin{figure}
\centering
\includegraphics[width=1\linewidth]{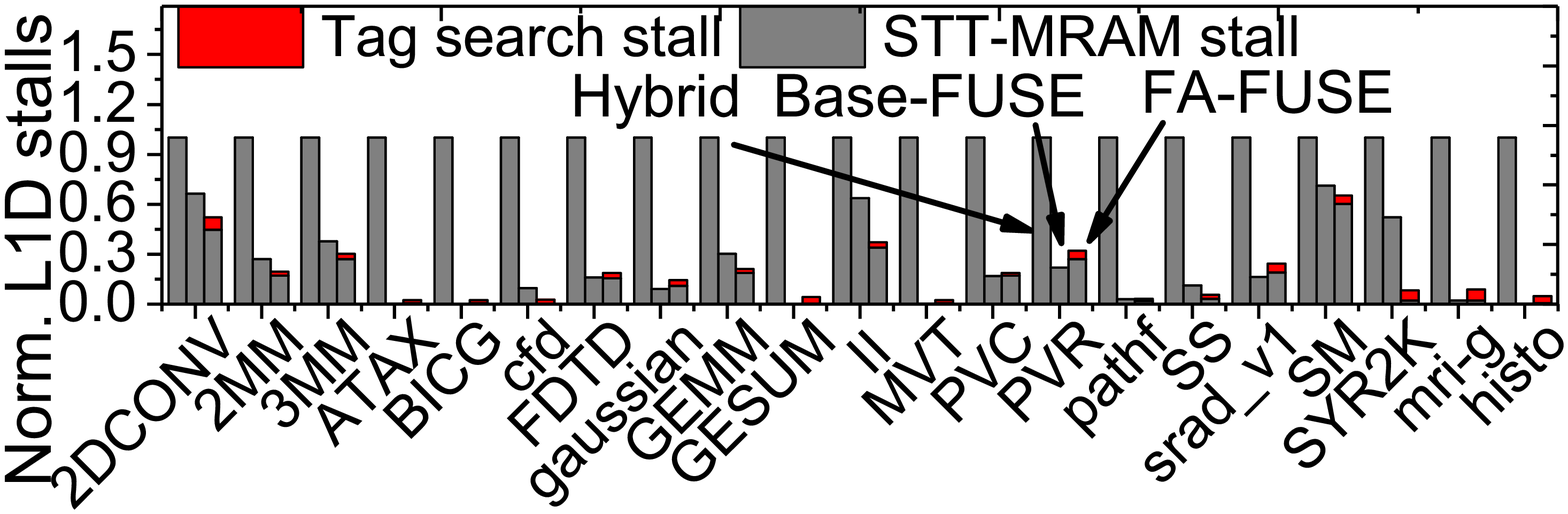}
\vspace{-20pt}
\caption{L1D cache stalls caused by STT-MRAM and tag searching.
\vspace{-5pt}
}
\vspace{-10pt}
\label{fig:L1_stall}
\end{figure}

\begin{figure}
\centering
\vspace{-5pt}
\includegraphics[width=1\linewidth]{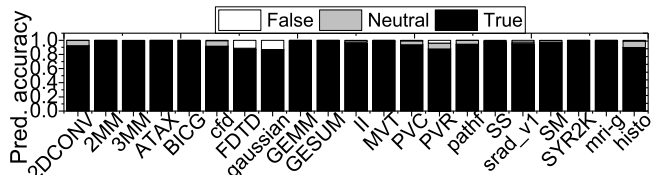}
\vspace{-20pt}
\caption{Read-level predictor accuracy.
}
\label{fig:Predictor_fig}
\end{figure}

\noindent \textbf{L1D cache stalls.} 
Figure \ref{fig:L1_stall} shows the number of stalls caused by STT-MRAM writes (STT-MRAM stall) and tag searching (tag search stall) in \texttt{Hybrid}, \texttt{Base-FUSE}, and \texttt{FA-FUSE} across all evaluated workloads. Note that the stalls are normalized to the STT-MRAM stalls of \texttt{Hybrid}. As shown in the figure, \texttt{Base-FUSE} can effectively reduce the number of L1D cache stalls by 78\%, compared to \texttt{Hybrid}, on average. This is because \texttt{Base-FUSE} enables SRAM bank and the associativity approximation logic unblocked, even though STT-MRAM is busy for serving data writes. On the other hand, \texttt{FA-FUSE} can reduce the number of L1D stalls by 18\%, compared to \texttt{Base-FUSE}. This is because the fully-associative STT-MRAM bank in \texttt{FA-FUSE} can keep more data requests which in turn reduces the number of cache misses in the L1D STT-MRAM bank. While the associativity approximation logic introduces the penalty of tag search stall in \texttt{FA-FUSE}, this overhead is only 3\% of STT-MRAM stalls in \texttt{Hybrid}.

\noindent \textbf{Read-level predictor.} We set up the read-level predictor based on empirical configurations of the entries and sets in sampler as well as the entries and the thresholds in prediction history table (cf., Table \ref{tab:system_config}). Figure \ref{fig:Predictor_fig} shows the prediction accuracy of our read-level predictor for each workload. Specifically, we mark a prediction as ``True", if data is predicted as WM and it experiences multiple writes before eviction or data is predicted as WORM/WORO and it experiences only singular write. In comparison, the prediction is marked as ``False", if data is predicted as WM but it experiences singular write before eviction, or data is predicted as WORM/WORO but it experiences multiple writes. Finally, the prediction is marked as ``Neutral", if the read-level predictor could not provide prediction. As shown in Figure \ref{fig:Predictor_fig}, our read-level predictor can give prediction accuracy of 95\%, on average, across all workloads, and the prediction accuracy decreases to 85\% in the worst case. Thanks to the accurate prediction, \texttt{Dy-FUSE} can significantly reduce the number of data migrations between SRAM and STT-MRAM, which addresses the overheads caused by STT-MRAM writes.

\begin{figure}
\centering
\includegraphics[width=1\linewidth]{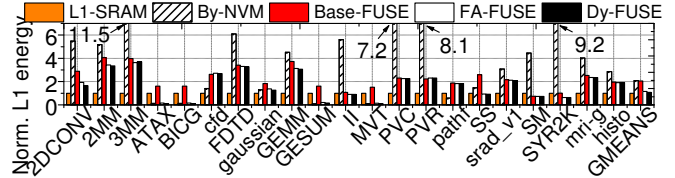}
\vspace{-20pt}
\caption{L1D energy normalized to \texttt{L1-SRAM}.
\vspace{-15pt}
}
\vspace{-5pt}
\label{fig:L1_eneg1}
\end{figure}

\noindent \textbf{L1D Energy.}  
Figure \ref{fig:L1_eneg1} plots the L1D cache energy of different cache configurations, which are normalized to \texttt{L1-SRAM}. As shown in the figure, \texttt{L1-SRAM} consumes the lowest energy in the compute-intensive workloads that exhibit low APKI (e.g., \emph{2DCONV}, \emph{2MM}, and \emph{3MM}). This is because the small-size SRAM cache consumes much lower read/write power than STT-MRAM of the same area size (cf. Table \ref{tab:system_config}). On the other hand, for the irregular workloads and data-intensive workloads (e.g., \emph{ATAX}, \emph{BICG}, and \emph{MVT}), \texttt{L1-SRAM} consumes 6x, 5.6x, and 8.4x higher energy higher than \texttt{By-NVM}, \texttt{FA-FUSE}, and \texttt{Dy-FUSE}. Since \texttt{L1-SRAM} requires much longer execution time due to the high L1D cache miss rate, it wastes a huge amount of energy due to SRAM's high leakage power.
Even though By-NVM can bypass deadwrite to avoid unnecessary write energy consumption in STT-MRAM, it still suffers from the low energy efficiency imposed by cache write updates or read misses. On the other hand, FUSE can accommodate most of the write updates and read misses in SRAM, which in turn reduces the write energy.
\texttt{FA-FUSE} and \texttt{Dy-FUSE} can reduce energy consumption by 16\% and 24\% compared to \texttt{By-NVM}, as they feature sufficient memory (SRAM) space to accommodate the writes.
\texttt{Dy-FUSE} can reduce the energy consumption by 7\%, compared to \texttt{FA-FUSE}. This is because \texttt{Dy-FUSE} employs a read-level predictor to accurately put WM data in SRAM and read-intensive data in STT-MRAM separately.

\begin{figure}
\centering
\subfloat[IPC normalized to \texttt{1/16}.]{\label{fig:SRAMratio1}\rotatebox{0}{\includegraphics[width=0.49\linewidth]{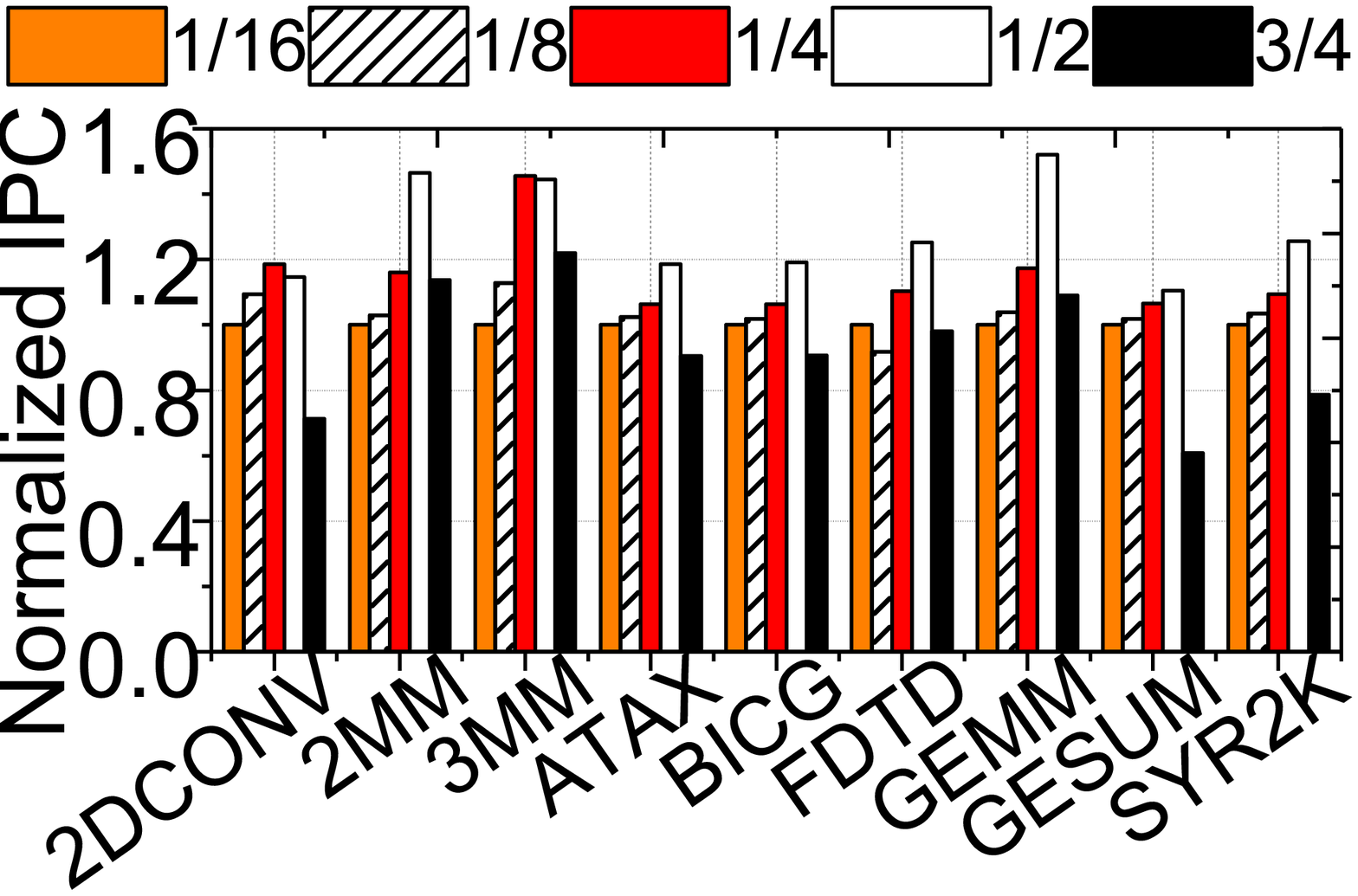}}}
\subfloat[L1D miss rate.]{\label{fig:SRAMratio2}\rotatebox{0}{\includegraphics[width=0.49\linewidth]{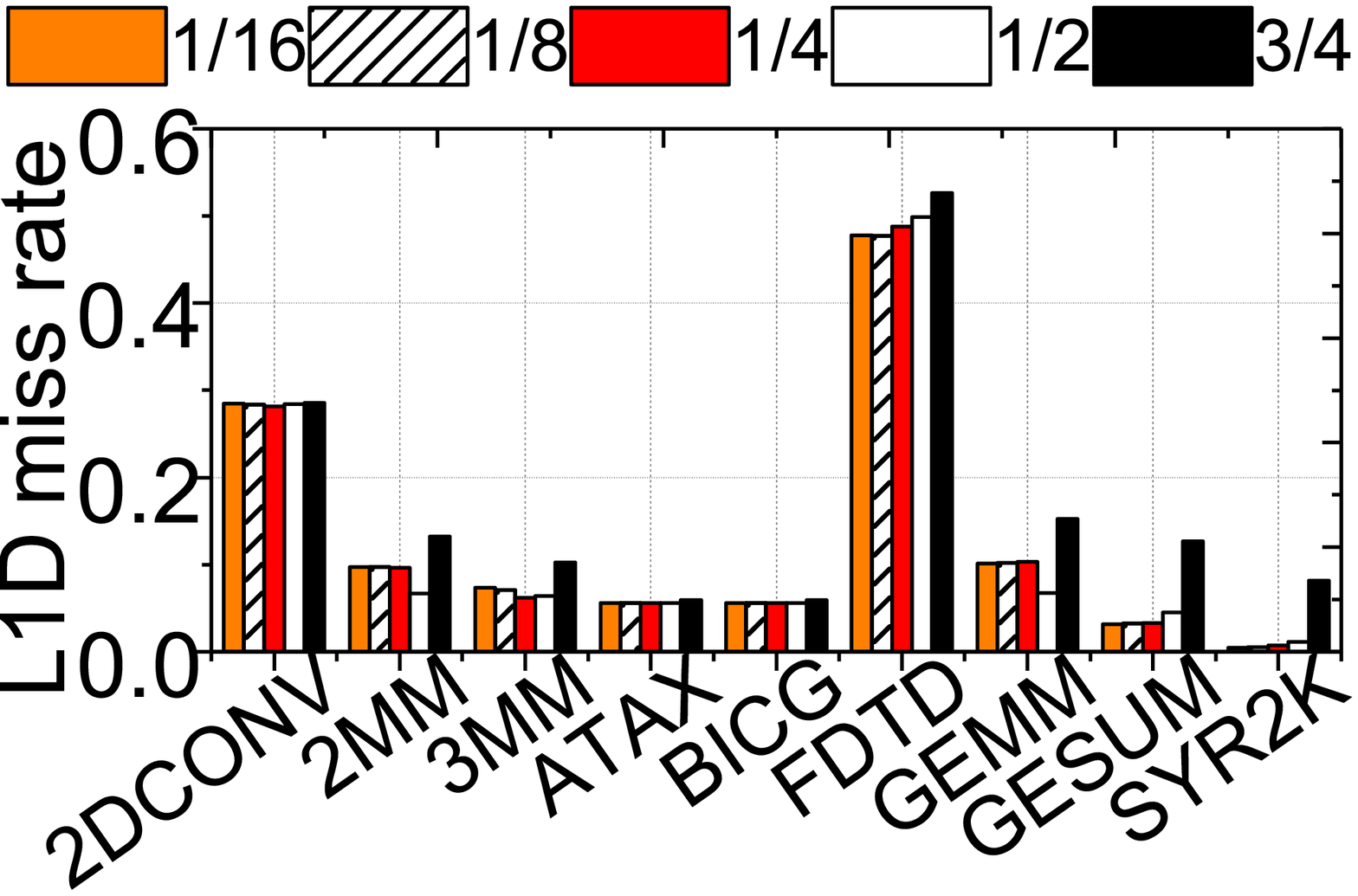}}}
\vspace{-5pt}
\caption{Performance comparison of different SRAM and STT-MRAM ratio in L1D cache.\vspace{-10pt}}
\label{fig:SRAMratio}
\end{figure}

\subsection{Sensitivity Study}
\noindent \textbf{SRAM-to-STT-MRAM ratios.} We evaluate various STT-MRAM/SRAM configurations to seek the optimal one that exhibits the best performance across all workloads we tested. Figure \ref{fig:SRAMratio} shows the IPC and L1D miss rates of different configurations which are normalized to ``1/16". Note that, \texttt{1/16} indicates that 1/16 of L1D cache space is allocated as SRAM bank and the remaining space is used by STT-MRAM. As shown in Figure \ref{fig:SRAMratio1}, \texttt{1/2} achieves the best performance among all the configurations that we evaluated. This is because, compared to \texttt{1/2}, allocating more space for SRAM (e.g., \texttt{3/4}) reduces the overall L1D cache capacity by 30\%, which in turn increases the cache miss rate compared to \texttt{1/2} (cf., Figure \ref{fig:SRAMratio2}) by 4\%, on average. Even though allocating more space for STT-MRAM (e.g., \texttt{1/16}, \texttt{1/8}, and \texttt{1/4}) can increase L1D cache space by 53\%, 45\%, and 30\%, compared to \texttt{1/2}, only 0.7\%, 0.68\%, and 0.51\% of L1D cache miss rate reduction can be observed in Figure \ref{fig:SRAMratio2}. This is because our associativity approximation logic can maximize the utilization of STT-MRAM cache, which reduces the demands for bigger cache capacity. However, since less space is allocated for SRAM in \texttt{1/16}, \texttt{1/8}, and \texttt{1/4}, the limited capacity of SRAM cannot accommodate all write-multiple (WM) data, which in turn increases the number of STT-MRAM writes and introduces a significant overhead imposed by write penalties of STT-MRAM. Based on the above evaluation results, we identify \texttt{1/2} as the optimal STT-MRAM/SRAM configuration.

\begin{figure}
\centering
\includegraphics[width=1\linewidth]{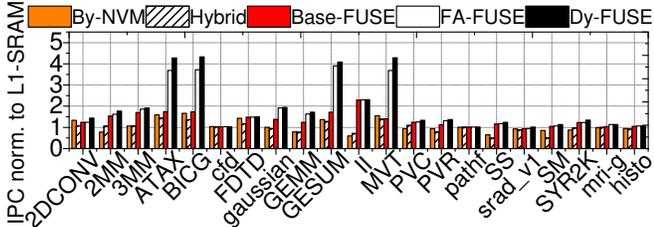}
\vspace{-20pt}
\caption{Performance in Volta GPU.
\vspace{-5pt}
}
\vspace{-15pt}
\label{fig:ipc_128KB}
\end{figure}

\noindent \textbf{Volta GPU.} By default, GPGPU-Sim simulator only provides configurations that mimic Fermi \cite{wittenbrink2011fermi} and Kepler \cite{keplernvidia} architectures. Consequently, we revised its GPU architecture model to model the latest Volta GPU \cite{voltanvidia}. Specifically, we increased the number of SMs to 84, modified the L2 cache size to 6 MB, and increased the memory bandwidth to 900 GB/s. As the L1 cache in Volta GPU is reconfigurable (ranging from 32 KB to 128 KB), we configured a 128 KB L1 cache to evaluate the impact of large L1 cache. Figure \ref{fig:ipc_128KB} gives the performance of \texttt{L1-SRAM}, \texttt{By-NVM}, \texttt{Hybrid}, and our FUSE caches in terms of IPC values under the tested workloads. All the results are normalized to \texttt{L1-SRAM}. As shown in the figure, it is beneficial to enlarge the L1D cache capacity further by employing STT-MRAM. Specifically, compared to \texttt{L1-SRAM}, \texttt{By-NVM} is able to accommodate more memory requests in its L1D cache, which in turn improves the overall performance by 57\% in the irregular workloads (e.g., \emph{ATAX} and \emph{GESUM}). While \texttt{By-NVM} achieves poor performance in the write-intensive workloads due to its long STT-MRAM write latency, thanks to the optimized architecture that fuses SRAM and STT-MRAM, the \texttt{Base-FUSE}, \texttt{FA-FUSE} and \texttt{Dy-FUSE} successfully mitigate the write penalty, which in turn improves the performance by 37\%, 71\%, and 82\%, compared to \texttt{By-NVM}. Overall, \texttt{Base-FUSE}, \texttt{FA-FUSE} and \texttt{Dy-FUSE} improve the performance by 35\%, 82\%, and 96\%, compared to \texttt{L1-SRAM}.

\begin{figure}
\centering
\def\subfigcapskip{0pt}
\subfloat[CBF hash functions.]{\label{fig:CBFtag}\rotatebox{0}{\includegraphics[width=0.47\linewidth]{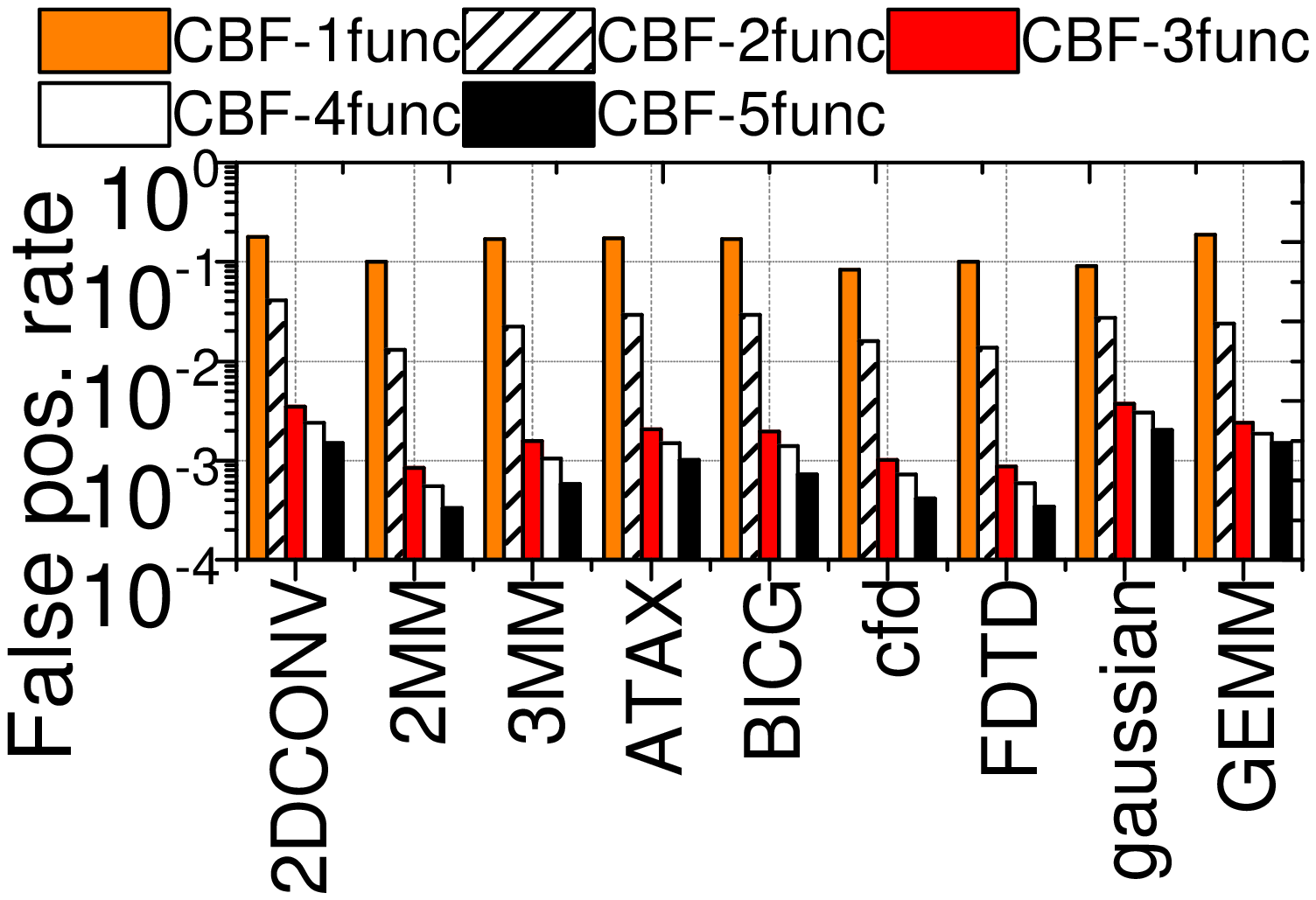}}}
\hspace{3pt}
\subfloat[CBF data set slots.]{\label{fig:hashfunctiontag}\rotatebox{0}{\includegraphics[width=0.48\linewidth]{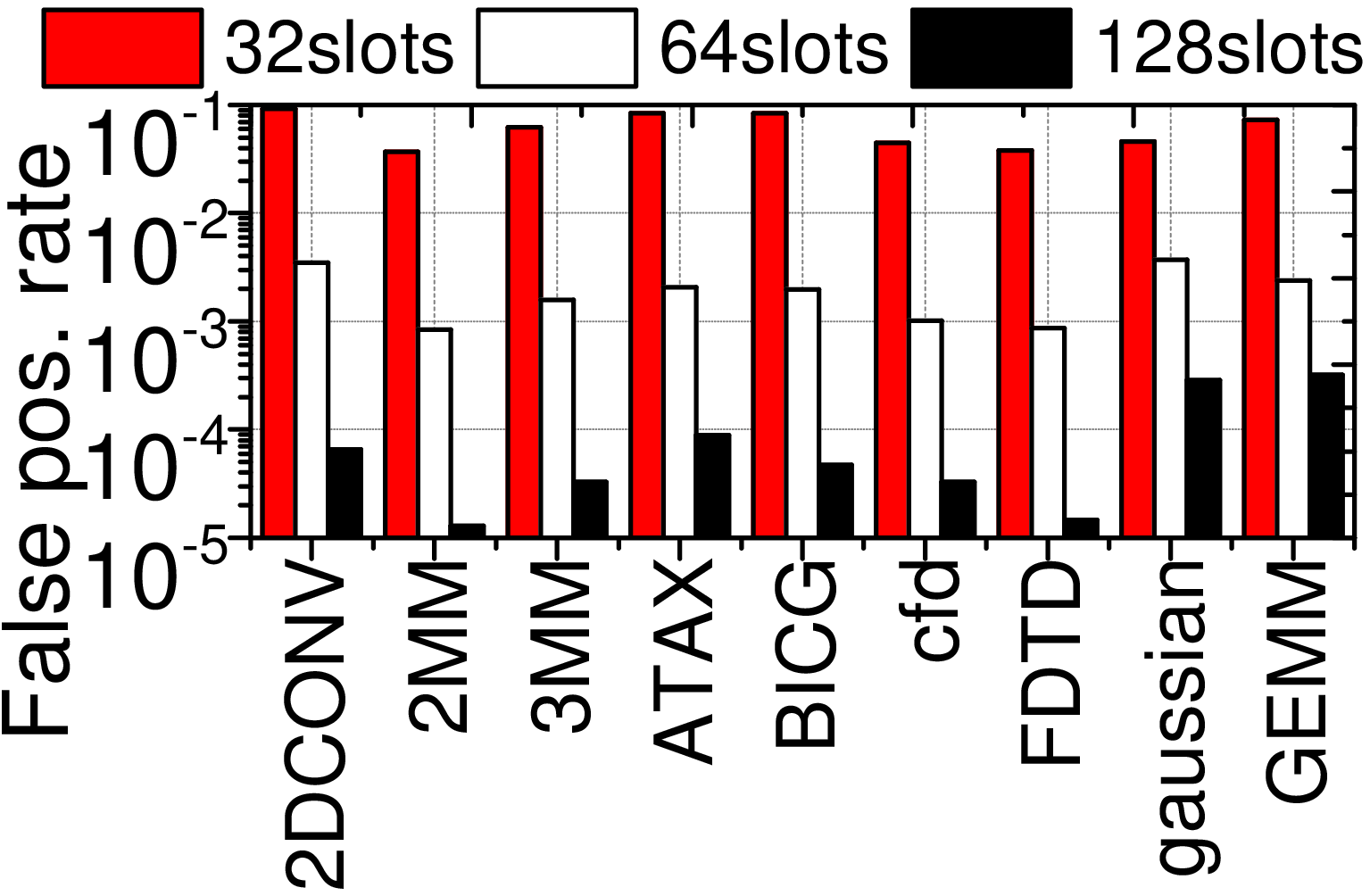}}}
\vspace{-5pt}
\caption{The false positive rate in CBF.\vspace{-10pt}}
\label{fig:CBFperf}
\vspace{-10pt}
\end{figure}

\noindent \textbf{False positives of CBF.} The performance of CBF is one of the factors that impact the efficiency of the associativity approximation logic and the overall system performance. Figure \ref{fig:CBFperf} illustrates the effects of employing a different number of hash functions and increasing the data set sizes on the false positive rate. \texttt{CBF-\underline{X}func} indicates employing \underline{X} hash functions and \texttt{\underline{X}slots} means each data set has \underline{X} slots. Increasing the number of hash functions effectively reduces the number of false positives (cf. Figure \ref{fig:CBFtag}). Specifically, \texttt{CBF-3func} with 3 hash functions reduces the number of false positives by 98.4\%, compared to \texttt{CBF-1func} with single hash function. While employing more hash functions can reduce the number of false positives further, the performance improvement would get saturated. The large data set space such as \texttt{128slots} and \texttt{64slots} reduces the number of false positives by 99\% and 97\%, respectively, compared to \texttt{32slots} (cf. Figure \ref{fig:hashfunctiontag}). This is because, a larger data set space reduces the possibility for hash functions to generate same hash key from different inputs. To minimize the impact of false positives, we select \texttt{128slots} as our CBF configuration.

\begin{table}
\centering
\resizebox{\columnwidth}{!}{
\begin{tabular}{|l|c|c|c|c|}
\hline
 & \textbf{Components} & \textbf{Transistors} & \textbf{Components} & \textbf{Transistors} \\ \hline
\multirow{3}{*}{\textbf{L1-SRAM}} & data array & 1,572,864 & write driver & 58,520 \\ \cline{2-5} 
 & tag array & 32,256 & comparator & 976 \\ \cline{2-5} 
 & sense amplifier & 66,880 & decoder & 1,124 \\ \hline
\multirow{5}{*}{\textbf{Dy-FUSE}} & data array & 1,572,864 & write driver & 45,980 \\ \cline{2-5} 
 & tag array & 43,776 & comparator & 1,458 \\ \cline{2-5} 
 & sense amplifier & 48,070 & decoder & 1,686 \\ \cline{2-5} 
 & NVM-CBF & 10,944 & swap buffer & 3,072 \\ \cline{2-5} 
 & request queue & 15,360 & read-level predic. & 2,320 \\ \hline
\end{tabular}}
\vspace{-5pt}
\caption{Area estimation of \texttt{L1-SRAM} and \texttt{Dy-FUSE}.\vspace{-5pt}\label{tbl:area}}
\vspace{-15pt}
\end{table}

\subsection{Area Overhead Analysis}
\label{sec:area_cost}
Table \ref{tbl:area} lists the area estimation of each component employed in a 32KB \texttt{L1-SRAM} and \texttt{Dy-FUSE} in terms of the number of transistors.

\noindent \textbf{SRAM-based cache.}
Considering that a 6T structure of an SRAM cell, the 32KB data array requires 1.5 million transistors. We assume each tag entry contains 19-bit tag, 1-bit valid bit, and 1-bit dirty bit. Thus, 32K transistors are required for a 32KB tag array. The sense amplifier of SRAM-based cache typically composes 8T sensing and 8T latch circuits to sense and hold 1-bit data, while the write driver consists of 14T to write 1-bit data. Overall, the SRAM-based cache needs 37K and 32K transistors to build its sensing amplifier and write driver, respectively. In our design, 4T comparison circuit can compare 1 tag bit; the total number of transistors to build comparators is around 976. Lastly, the procedure of address decoding can be divided into three stages: predecoding, combination, and wordline driving. Among them, the predecoding stage uses a couple of 2-4 and 3-8 decoders, the combination stage employs a NOR gate for each wordline, and the wordline driving stage utilizes tri-state inverters to drive each wordline. Considering all, the required number of transistors for address decoding is 1124.

\noindent \textbf{\texttt{Dy-FUSE}.} 
Thanks to the serialization of our tag array and data array accesses in the STT-MRAM, FUSE architecture is able to reduce the space allocated for sensing amplifier and write driver in the SRAM-based cache, and leverages the free space to accommodate 4 extra components: two 128-bit sensing amplifiers, two decoders, and 128 sets of NVM-CBFs, each containing 64 2-bit STT-MRAM based counters (4 transistors and 2 MTJ for each counter); a swap buffer with 3 entries, each entry requiring 1024 transistors; a request queue with 16 entries, each entry requiring 960 transistors; a sampler and a prediction table within a read-level predictor, require 648 transistors and 1672 transistors, respectively. Our synthesis analysis reports that Dy-FUSE exceeds the L1D cache area by less than 0.7\%.

\section{Discussion}
\label{sec:discuss}
Embedded DRAM (eDRAM) and STT-MRAM are promising memory technologies as CPU/GPU cache. While eDRAM can provide higher cache density than SRAM, its feature size (60-100 $F^{2}$) is much bigger than that of STT-MRAM (6-50 $F^{2}$). As FUSE prefers large capacities, STT-MRAM is selected as candidate. STT-MRAM also exhibits better performance and higher power-efficiency than eDRAM, as eDRAM incurs extra time and power to refresh its cells in every 40 us.

However, STT-MRAM suffers from long write latency. Prior work \cite{wang2014coherent,tan2015soft} proposed to leverage SRAM to hide this write penalty and explored the feasibility of architecting CPU L1 cache and GPU register file with a SRAM/STT-MRAM hybrid. Fabricating SRAM/STT-MRAM hybrid in the same die is also proven to be feasible. While the commercial prototype of the hybrid SRAM and STT-MRAM cache is currently not available in the market, \cite{durlam20031} presents a research sample which integrates STT-MRAM cells into CMOS circuit with copper interconnect technology.

Unlike CPU, GPU L1D cache does not support any cache coherence protocol in hardware. Instead, it adopts a weak consistency model, which relies on the synchronization between the kernels as well as the synchronization primitives in user programs to maintain data consistency. While prior work \cite{jia2012characterizing,li2015locality} assumes that CUDA cores write through dirty L1D cache blocks to the shared L2 cache to ensure data consistency, L1D cache actually is implemented as write-back cache with the support of synchronization. Thus, in this paper, we adopt a write-back policy in L1D cache.

To allocate more L1D cache space for the data-intensive workloads, academia and industry adopt a unified on-chip memory space consisting of L1D cache and shared memory (as well as register files). When allocating more memory space as L1D cache, it has to reduce the capacity of shared memory and register files. This in turn can hurt the performance of data-intensive workloads, which have high demands for large shared memory and register files. In contrast, FUSE increases the L1D cache size by leveraging larger STT-MRAM bank, which does not sacrifice the size of shared memory and register files. In addition, FUSE also simplifies the circuit design of the unified memory space.

\section{Related Work}
\label{sec:relatedwork}
\noindent \textbf{Cache bypassing and warp throttling.} There are many prior studies oriented towards eliminating GPU L1D cache thrashing. For example, \cite{chen2014adaptive, ausavarungnirun2015exploiting, li2015priority, xie2015coordinated, xie2013efficient} propose to bypass part of data requests to the off-chip memory in an attempt to protect data blocks in L1D cache from early eviction. However, these bypass strategies can degrade the overall performance and make energy efficiency worse, since they typically introduce more off-chip access overheads. In addition, it is challenging to employ them in an STT-MRAM cache, as they are unaware of write accesses, which can introduce long write delays. Rogers et al \cite{rogers2012cache} propose to throttle the number of warps that compete for L1D cache so as to minimize cache thrashing. However, doing so can harm the overall performance due to the reduced thread-level parallelism. 
In contrast, FUSE maximizes the number of active threads, while minimizing the number of off-chip memory accesses by employing an NVM-based full-associative L1D cache to accommodate as many memory requests as possible.

\noindent \textbf{Hybrid caches.} Many prior studies \cite{khoshavi2016read, zhan2016oscar, wang2014coherent, rodriguez2015volatile} have successfully orchestrated STT-MRAM in CPU caches by being aware of the access patterns exhibited by the CPU. However, they, unfortunately, cannot be directly applied to GPGPU, as the SIMT execution on GPU generates different access patterns from CPU. In particular, the read-level predictor in FUSE is tailored to GPU, which can speculate access patterns generated by SIMT execution.
Sun et al. and Choi et al. \cite{sun2009novel, coi2017novel} detail the design of heterogeneous SRAM/STT-MRAM architectures. More specifically, while \cite{coi2017novel} proposes to architect an SRAM-based tag array and an STT-MRAM-based data array to leverage the unique advantages of each, the proposed hybrid cache architecture requires a large cache capacity and is sensitive to I/O access patterns, which may not be beneficial for the small-sized GPU L1D cache. In comparison, \cite{sun2009novel} propose a new cache layout for hybrid SRAM and STT-MRAM in CPU shared L2 NUCA cache. However, this design may not blend well with the private L1D cache in GPUs. Compared to prior work, FUSE customizes its hybrid cache architecture to work with the GPU L1D cache by carefully considering area constraints, cache layout, and peripheral circuits.

\noindent \textbf{STT-MRAM integration in GPUs.} There also exist prior efforts that try to incorporate NVM in a GPU memory hierarchy. Samavatian et al. \cite{samavatian2014efficient} presents a new L2 cache that integrates low-retention STT-MRAM and long-retention STT-MRAM arrays. To reduce the write penalty of STT-MRAM, \cite{samavatian2014efficient} proposes to store write-intensive data in low-retention STT-MRAM and store read-only data in high-retention STT-MRAM. Compared to \cite{samavatian2014efficient}, we propose to fuse L1 SRAM with STT-MRAM, which can eliminate the refresh overhead of low-retention STT-MRAM. In addition, FUSE employs a new approximate full-associative cache architecture and introduces a highly-accurate read-level predictor based on program-counter. Wang and Xie \cite{wang2015write} propose an STT-MRAM based GPU register file architecture. It addresses the write penalty of STT-MRAM by enabling simultaneous reads and writes in the same bank, and by adding an SRAM buffer to accommodate repeated writes. However, such techniques cannot be applied to the L1D cache, as it has a different architecture compared to register file, and the repeated writes on the same data are not observed in L1D cache.

\noindent \textbf{Full associativity.} Jouppi \cite{jouppi1990improving} proposes a full-associative victim cache to decrease the number of conflict misses caused by direct-mapped caches in CPUs. However, this approach cannot aid to improve cache hit rate in GPU, as the limited capacity of victim cache (e.g., 16 cache lines) makes it hard to accommodate all cache misses from a massive number of threads (more than thousand). In contrast, FUSE leverages an associativity approximation logic to configure a full-associative L1D cache, and this provides sufficient capacity to reduce as many cache misses as possible.




\ignore{
Increasing L1D cache capacity can be a thrashing-resistant solution. However, it is not practical to allocate large space for L1D cache in already crowded on-chip memory. \cite{chen2014adaptive} proposes to bypass part of data requests to underlying L2 cache in attempt to protect cache lines from early eviction. However, the bypass strategy can degrade the overall system performance and make energy efficiency worse since it can introduce many off-chip access overheads. \cite{rogers2012cache} proposes to throttle the number of warps to reduce the competition for L1D cache. However, it can seriously degrade performance due to reduced thread-level parallelism.
}

\section{Conclusion}
\label{sec:conclusion}
In this work, we proposed FUSE, a novel heterogeneous GPU cache that effectively increases L1D cache capacity by fusing STT-MRAM into on-chip L1D cache. 
FUSE can improve the overall GPU performance by reducing the overheads imposed by off-chip memory accesses, while efficiently mitigating the write penalty of STT-MRAM. 
Our extensive evaluations indicate that FUSE exhibits 217\% better performance compared to a traditional SRAM-based approach and saves about 53\% energy over it. 

\section{Acknowledgement}
This research is mainly supported by NRF 2016R1C1B2015312, DOE
DEAC02-05CH11231, IITP-2018-2017-0-01015, NRF 2015M3C4A7065645, Yonsei
Future Research Grant (2017-22-0105) and MemRay grant (2015-11-1731).
M. Kandemir is supported in part by grants by NSF grants 1822923, 1439021,
1629915, 1626251, 1629129, 1763681, 1526750 and 1439057.
Myoungsoo Jung is the corresponding author who has ownership of this work.

{
\bibliographystyle{IEEEtran}
\bibliography{fuse}
}
\end{document}